\documentclass[aps,prd,nofootinbib,twocolumn,floatfix,superscriptaddress,eqsecnum]{revtex4}
\usepackage{dcolumn}
\usepackage{times,mathptm}
\usepackage{subfigure}
\usepackage{graphicx,psfrag,pifont,color,setspace}
\usepackage{amsfonts, amsthm,bm,amsmath,amssymb,marvosym}
\usepackage[automark]{scrpage2}
\newcommand{\beq}{\begin{equation}}
\newcommand{\eeq}{\end{equation}}
\newcommand{\bea}{\begin{eqnarray}}
\newcommand{\eea}{\end{eqnarray}}
\newcommand{\cb}{$\chi SB$}

\renewcommand{\l}{\lambda}

\renewcommand{\b}{\beta}

\newcommand{\m}{\mu}

\newcommand{\s}{\sigma}

\newcommand{\rf}[1]{(\ref{#1})}
\newcommand{\ra}{\rightarrow}

\begin{document}

\title{Center Vortices and the Dirac Spectrum}
\author{R. H\"ollwieser}
\affiliation{Atomic Institute, Technical University of Vienna, Wiedner Hauptstr.\ 8-10, A-1040 Vienna, Austria}

\author{M. Faber}
\affiliation{Atomic Institute, Technical University of Vienna, Wiedner Hauptstr.\ 8-10, A-1040 Vienna, Austria}

\author{J. Greensite}
\affiliation{Physics and Astronomy Dept., San Francisco State
University, San Francisco, CA~94132, USA}

\author{U.M. Heller}
\affiliation{American Physical Society, One Research Road, Box 9000,
Ridge, NY 11961-9000, USA}

\author{{\v S}. Olejn\'{\i}k}
\affiliation{Institute of Physics, Slovak Academy
of Sciences, SK--845 11 Bratislava, Slovakia}

\date{\today}
\begin{abstract}
We study correlations between center vortices and the low-lying eigenmodes of the Dirac operator, in both the overlap and asqtad formulations. In particular we address a puzzle raised some years ago by Gattnar et al.\ [Nucl. Phys. B 716, 105 (2005)], who noted that the low-lying Dirac eigenmodes required for chiral symmetry breaking do not appear to be present in center-projected configurations.  We show that the low-lying modes are in fact present in the staggered (asqtad) formulation, but not in the overlap and ``chirally improved'' formulations, and suggest a reason for this difference. We also confirm and extend the results of Kovalenko et al.\ [Phys. Lett. B 648, 383 (2007)], showing  that there is a correlation between center vortex locations, and the scalar density of low-lying Dirac eigenmodes derived from unprojected configurations.  This correlation is strongest at points which are associated, in the vortex picture, with non-vanishing topological charge density, such as vortex intersection and ``writhing'' points. We present supporting evidence that the lowest Dirac eigenmodes, in both asqtad and overlap formulations, have their largest concentrations in point-like regions, rather than on submanifolds of higher dimensionality.

\end{abstract}

\maketitle

%
%
\section{Introduction}\label{Intro}

There are two reasons $-$ ``theoretical'' and ``experimental'' $-$ to believe that center vortices are the dominant feature of the vacuum state of pure $SU(N)$ gauge theories at large scales. The theoretical reason is the simple fact that the asymptotic string tension of static color charge sources depends only on the $N$-ality of the color charges. This fact can be understood as arising from color charge screening by gluons, which is an explanation in terms of particle excitations. But there should also be a corresponding ``field'' explanation of $N$-ality dependence, in terms of gauge field configurations which dominate the path integral at large scales, and the center vortex scenario is the only explanation of this type which is known. The ``experimental'' reason for believing in vortex dominance is the wealth of numerical evidence in its favor, which was summarized a few years ago in Ref.\ \cite{review}. A key feature of this evidence is the strong correlation between vortex location, determined by center gauge-fixing and projection methods, and gauge-invariant observables, such as action density and the phase of Wilson loops.

Center vortices were originally introduced to explain confinement, but a force strong enough to confine quarks is also generally expected to break chiral symmetry \cite{Casher}.\footnote{The converse, of course, is not true.  It is possible to have chiral symmetry breaking, as in the Nambu--Jona-Lasinio model, without having confinement.} According to the Banks-Casher analysis \cite{BC}, chiral symmetry breaking (\cb) is necessarily associated with a finite density of near-zero eigenmodes of the chiral-invariant Dirac operator, so we would of course expect this to be true for the Dirac spectrum evaluated in an ensemble of center-projected lattice configurations, which are known to be confining.

Several years ago, however, Gattnar et al.\ \cite{Gattnar} reported a puzzling result.  These authors computed the low-lying eigenvalue spectrum of a chirally-improved version of the Dirac operator due to Gattringer \cite{Gattringer}, which approximates Ginsparg-Wilson fermions. A dense set of near-zero eigenvalues was found for unmodified lattice configurations, as expected, and a large gap in the eigenvalue spectrum, centered at eigenvalue $\l=0$, opened up in the spectrum when center vortices were removed.  This gap was also expected, since the vortex-removed configurations are not confining, and it has been known for a long time that $\langle\overline{\psi}\psi \rangle \ra 0$ in vortex-removed configurations \cite{dFE}. The puzzle was that an even larger gap in the spectrum was found for center-projected configurations, which contain only thin vortex excitations, and which \emph{are} confining. Now if \cb ~was present in center-projected (or ``vortex-only'') configurations, and not in vortex-removed configurations, we would conclude that vortices explain \cb ~as well as confinement. If, on the other hand, \cb ~occured in vortex-removed configurations, but was absent in vortex-only configurations, we would then have to conclude that vortices are not especially relevant to \cb ~(although we would then like to understand how confinement can coexist with unbroken chiral symmetry).  But the finding that \cb , or, to be more precise, near-zero modes, are absent both in vortex-removed \emph{and} in vortex-only lattices was unexpected, and it poses a challenge to interpretation.  

The question which comes to mind is whether the large gap found by Gattnar et al.\ is related to the way in which chiral symmetry is realized on the lattice. The Casher argument \cite{Casher} that confinement implies \cb ~is based on the usual $SU(N_f)_L \times SU(N_f)_R$ symmetry of the continuum theory with massless fermions. However, the chirally-improved Dirac operator only approximates this symmetry for gauge-field configurations which vary smoothly at the lattice scale.  Center-projected configurations are not even close to smooth; plaquette variables make a sudden transition from the trivial center element outside the thin vortex, to a non-trivial center element inside. The chirally-improved Dirac operator is not necessarily chirally symmetric, even approximately, in such backgrounds. In the absence of a symmetry, there is no reason to expect spontaneous symmetry breaking.  If this fact explains why there is a gap in the eigenvalue spectrum of the chirally-improved operator, then it is reasonable to also expect a gap in the spectrum of the overlap operator~\cite{overlap}, when evaluated on center-projected configurations. Of course the overlap operator, in contrast to the chirally-improved operator, \emph{does} have an exact global symmetry, but the symmetry transformations are gauge-field dependent \cite{Luscher}, and only approximate the $SU(N_f)_L \times SU(N_f)_R$ chiral symmetry transformations of the continuum theory for configurations which vary slowly at the scale of the lattice spacing.  While this smoothness condition is expected in the continuum limit, it is never the case for center-projected configurations, and the Casher argument relating confinement to \cb ~need not apply.   

On the other hand, the Lagrangian for staggered fermions (and their asqtad cousins~\cite{ASQTAD}) is known to be invariant under a subgroup of the usual chiral symmetry, irrespective of the smoothness of the gauge-field background. If the puzzling gap in the Dirac eigenvalue spectrum found by Gattnar et al.\  is a consequence of the roughness of center-projected lattices, then we might expect this gap to disappear in the spectrum of the staggered or asqtad Dirac operators. Indeed, there is already a relevant result in ref.\ \cite{AdFE}, which reported that $\langle \overline{\psi} \psi \rangle > 0$ for staggered fermions on a center-projected lattice.  Likewise, suppose we somehow ``soften'' the center-projection procedure to make the center-projected configurations smoother, and evaluate the spectrum of the overlap operator on these smoothed vortex configurations. Then, if the roughness of thin vortices is the problem for the overlap formulation, the eigenvalue gap should go away for suitably smoothed, but still confining, vortex configurations. 
  
In section 2 we will report our results for the spectrum of the overlap and asqtad Dirac operators, when evaluated on normal, vortex-only (i.e.\ center-projected), and vortex-removed lattices.\footnote{Similar results for the overlap spectrum on vortex-removed lattices have been obtained
previously by Gubarev et al.\ \cite{morozov} and by Bornyakov et al.\ \cite{Borny}.} Those results support the view that center vortices alone can induce both confinement \emph{and}  chiral symmetry breaking. We go on in section 3, following the earlier work by Kovalenko et al.\ \cite{ITEP}, to report on other correlations between center-vortex location, and the density distribution of low-lying Dirac eigenmodes in overlap and asqtad formulations. These correlations are an important test of the picture advocated by Engelhardt and Reinhardt \cite{ER}, in which topological charge is concentrated at points where vortices either intersect, or twist about themselves (``writhe'') in a certain way, and Dirac zero modes are concentrated where the topological charge density is large. If topological charge is concentrated in point-like regions, as is the case in the vortex picture, and if zero (and near-zero) modes are concentrated in regions of high topological charge density, then one would expect that the eigenmode densities of low-lying eigenmodes would be peaked in point-like regions. In section 4 we provide some supporting evidence for this type of concentration. Our conclusions are found in Section 5.  In an Appendix we review details of the tadpole-improved L\"uscher-Weisz gauge action,  and report on some necessary checks of vortex location via center projection, in numerical simulations of this lattice action.
  
\section{Thin Vortices and Near-Zero Modes}\label{sec:vortices}

We begin with some preliminary information. Throughout this article we work with lattices generated by lattice Monte Carlo simulation of the tadpole improved L\"uscher-Weisz pure-gauge action, mainly at coupling $\b_{LW}=3.3$ (lattice spacing $a=0.15$ fm) for the $SU(2)$ gauge group. The locations of center vortices are identified as usual by mapping the $SU(2)$ lattice to a $Z_2$ lattice which contains, by definition, only thin vortex  excitations.  The mapping is carried out by fixing the lattice to the direct maximal center gauge, which is equivalent to Landau gauge in the adjoint representation, and which maximizes the squared trace of link variables. The gauge-fixing procedure is the over-relaxation method.\footnote{In this study we perform 150 gauge fixing iterations for five gauge copies, and for the best copy continue with the gauge fixing for a further 400 iterations. It is likely, however, that the selection of the best copy out of five copies is inessential, as the results are hardly distinguishable from a random choice of copy, as noted in the Appendix.}  The mapping to link variables on the center-projected (or ``vortex-only'') lattice, for the $SU(2)$ gauge group, is given by
\beq
 U_\m(x) \ra Z_\m(x) = \mbox{signTr}\Bigl[U_\m(x)\Bigr]
\label{project}
\eeq
and the link variables $U'$ on the vortex-removed lattice are defined as 
\beq
 U'_\m(x) = Z_\m(x) U_\m(x)
\eeq 
The claim is that the thin vortices of the center-projected lattice lie somewhere in the middle of thick center vortices (thickness $\approx 1$ fm) on the unprojected lattice, and that thick center vortices are responsible for the area-law falloff of large fundamental Wilson loops, as well as for the $N$-ality dependence of higher-representation string tensions in $SU(N)$. The justifications for these claims, obtained from Monte Carlo simulation of the Wilson action, were reviewed some years ago in Ref.\ \cite{review}. Similar tests in the case of the L\"uscher-Weisz action are reported in the Appendix. Given that thin vortices locate thick vortices, then in the ``vortex-removed'' lattice we are really inserting a thin vortex somewhere in the middle of a thick vortex; the effect is to cancel out the field of the thick vortex at large distances.

\begin{figure*}[t!]
  \centering
  \psfrag{ReL}{Re $\lambda$}
  \psfrag{ImL}{Im $\lambda$}
  \includegraphics[scale=0.80]{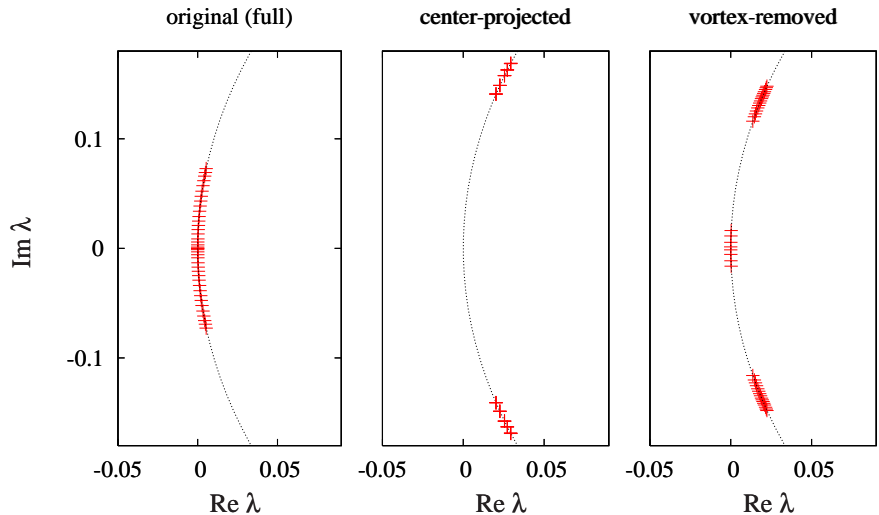}
  \caption{The first twenty overlap Dirac eigenvalue pairs on the Ginsparg-Wilson circle for a $16^4$ lattice, periodic boundary conditions at $\b_{LW}=3.3$. The center-projected configurations show a four-fold degeneracy.}\label{fig:ovlevs}
\end{figure*}

\begin{figure*}[]
  \centering
  \psfrag{ReL}{Re $\lambda$}
  \psfrag{ImL}{Im $\lambda$}
  \includegraphics[scale=0.80]{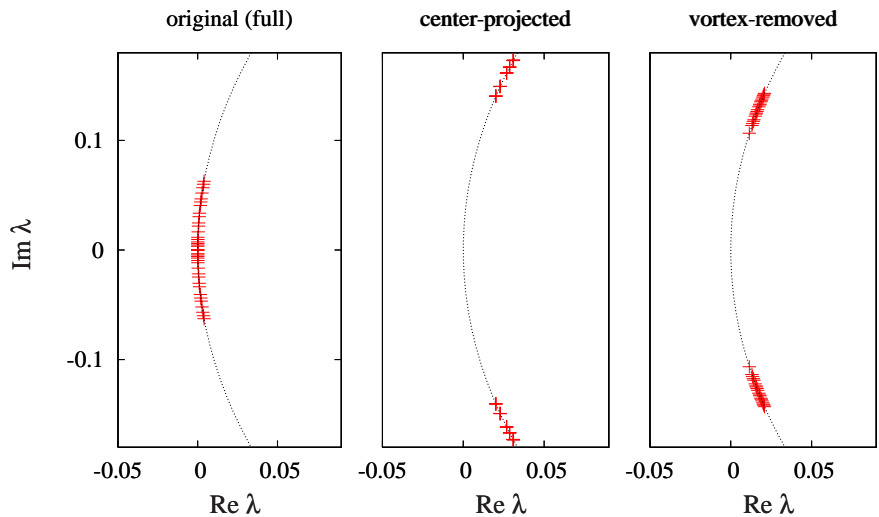}
  \caption{The first twenty overlap Dirac eigenvalue pairs on the Ginsparg-Wilson circle for a $16^4$ lattice, antiperiodic boundary conditions at $\b_{LW}=3.3$. The zero modes in vortex-removed configurations disappear.}\label{fig:apovlevs}
\end{figure*}

\begin{figure*}[]
  \centering
  \psfrag{ReL}{Re $\lambda$}
  \psfrag{ImL}{Im $\lambda$}
  \includegraphics[scale=0.80]{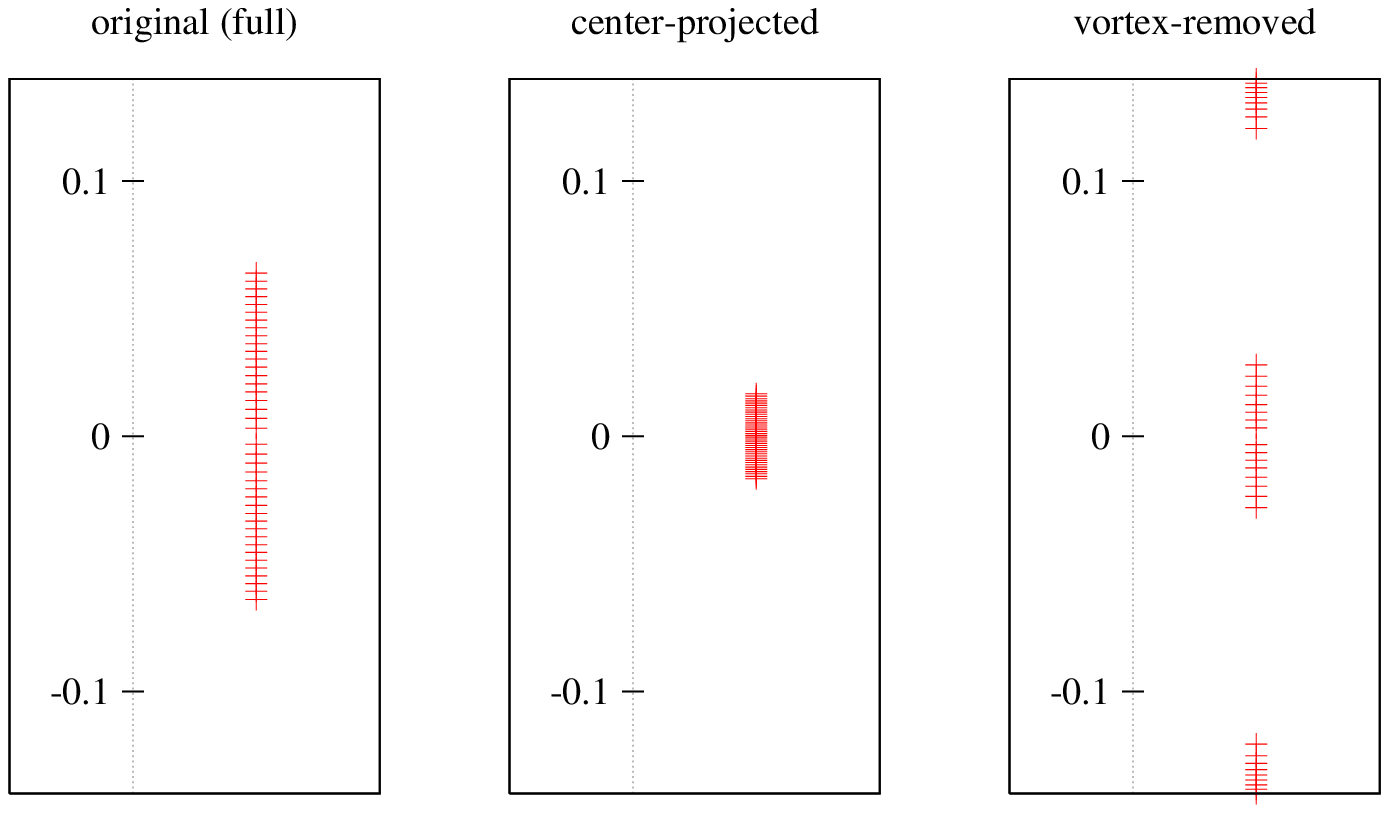}
  \caption{The first twenty asqtad Dirac eigenvalue pairs from a $16^4$ lattice, periodic boundary conditions at $\b_{LW}=3.3$.}\label{fig:stevs}
\end{figure*}

\begin{figure*}[]
  \centering
  \psfrag{ReL}{Re $\lambda$}
  \psfrag{ImL}{Im $\lambda$}
  \includegraphics[scale=0.80]{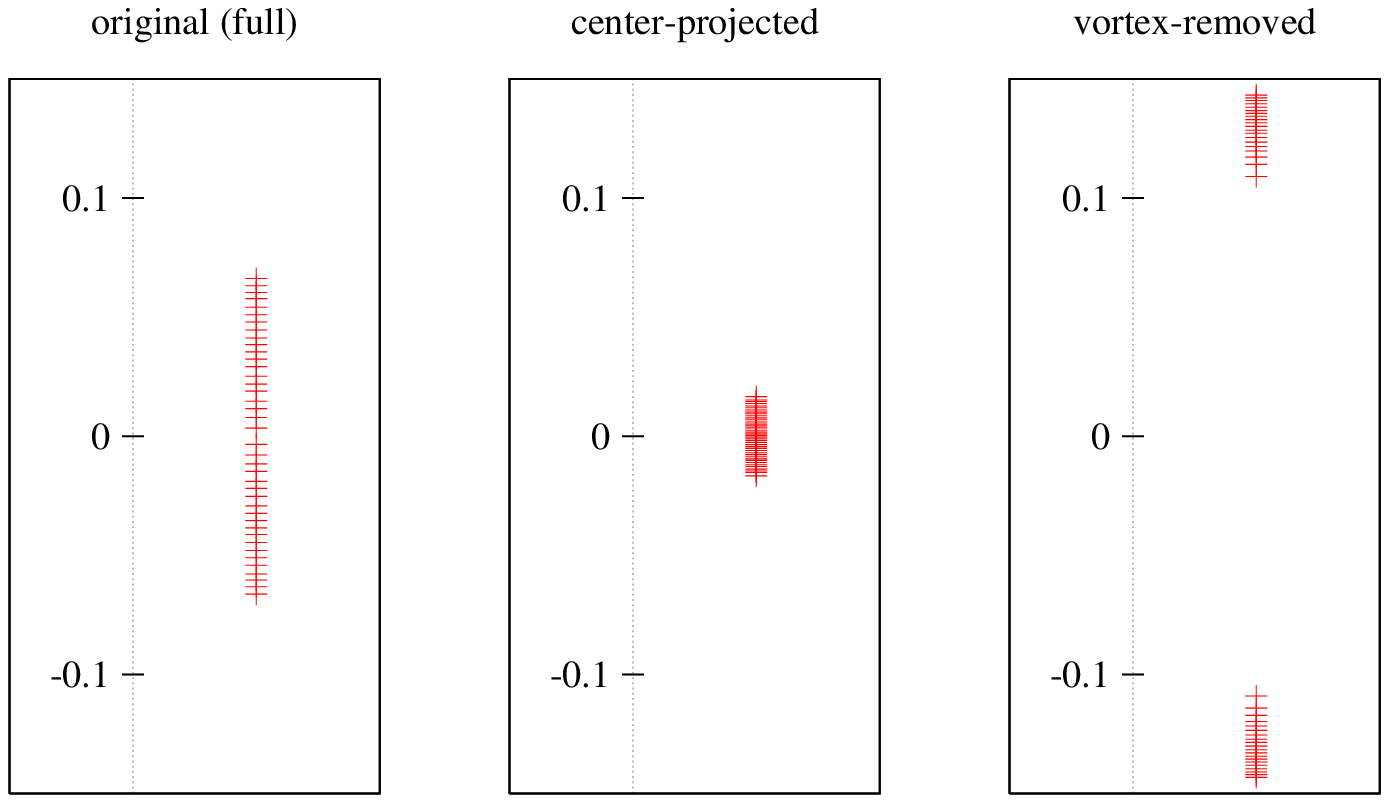}
  \caption{The first twenty asqtad Dirac eigenvalue pairs from a $16^4$ lattice, antiperiodic boundary conditions at $\b_{LW}=3.3$.}\label{fig:apstevs}
\end{figure*}

In Fig.\ \ref{fig:ovlevs} we display the twenty lowest-lying complex conjugate eigenvalue pairs of the overlap Dirac operator, on a $16^4$ lattice at $\b_{LW}=3.3$. Results are displayed for eigenmodes obtained on the original, center-projected and vortex-removed lattices. We observe the same phenomenon already reported by Gattnar et al.\ for the spectrum of the chirally-improved Dirac operator:  a large gap has opened in both the vortex-removed \emph{and} the vortex-only spectra. We note here that the gap in vortex-removed overlap spectrum was found previously, and discussed
in some detail, by Bornyakov et al. \cite{Borny}, who simulated a tadpole-improved Symanzik lattice action.
Looking more closely at the spectra, we see that there only appear to be five eigenvalue pairs (out of twenty) in the center-projected case, indicating a four-fold degeneracy when the overlap operator is applied to $Z_2$ lattice configurations.  This factor of four has the following origin:  In the first place, when link variables are simply plus or minus the $2\times 2$ identity matrix, the two colors decouple, and we have a factor of two degeneracy.
Secondly, whenever the link variables are real and the Dirac operator has the Wilson or overlap (but not staggered) form, the eigenvalue 
equation $D \psi_n = \l_n \psi_n$ is invariant under charge conjugation.  Thus, if $\psi_n$ is an
eigenstate with eigenvalue $\l_n$, then $C^{-1} \psi_n^*$ is also an eigenstate, with the same eigenvalue \cite{Leutwyler:1992yt}.  This gives
another factor of two, resulting in an overall four-fold degeneracy.  In the vortex-removed case, one does observe four near-zero modes of each chirality, but these can be interpreted as a remnant of the exact zero modes of the free theory that are associated with a periodic lattice.  These near-zero modes of the vortex-removed lattice are irrelevant to \cb, and disappear when we impose antiperiodic boundary conditions, as shown in Fig.\ \ref{fig:apovlevs}.      

In the Introduction we speculated that the reason for the large gap in the vortex-only case was connected with the lack of smoothness of center-projected lattices. In that case the exact symmetry of the overlap operator is strongly field-dependent, and does not really approximate the chiral symmetry of the continuum theory. Staggered and asqtad fermions, on the other hand, preserve a subgroup of the usual continuum $SU(N_f)_L \times SU(N_f)_R$ symmetry, irrespective of the smoothness of the configuration, and by the Casher argument \cite{Casher} one would expect this remaining symmetry to be spontaneously broken, on the lattice, by any ensemble of gauge configurations with the confinement property.  Then, according to the Banks-Casher relation, there should not be any gap in the vortex-only eigenvalue spectrum.

\begin{figure*}[t!]
  \centering
  \psfrag{ReL}{Re $\lambda$}
  \psfrag{ImL}{Im $\lambda$}
  \includegraphics[scale=0.9]{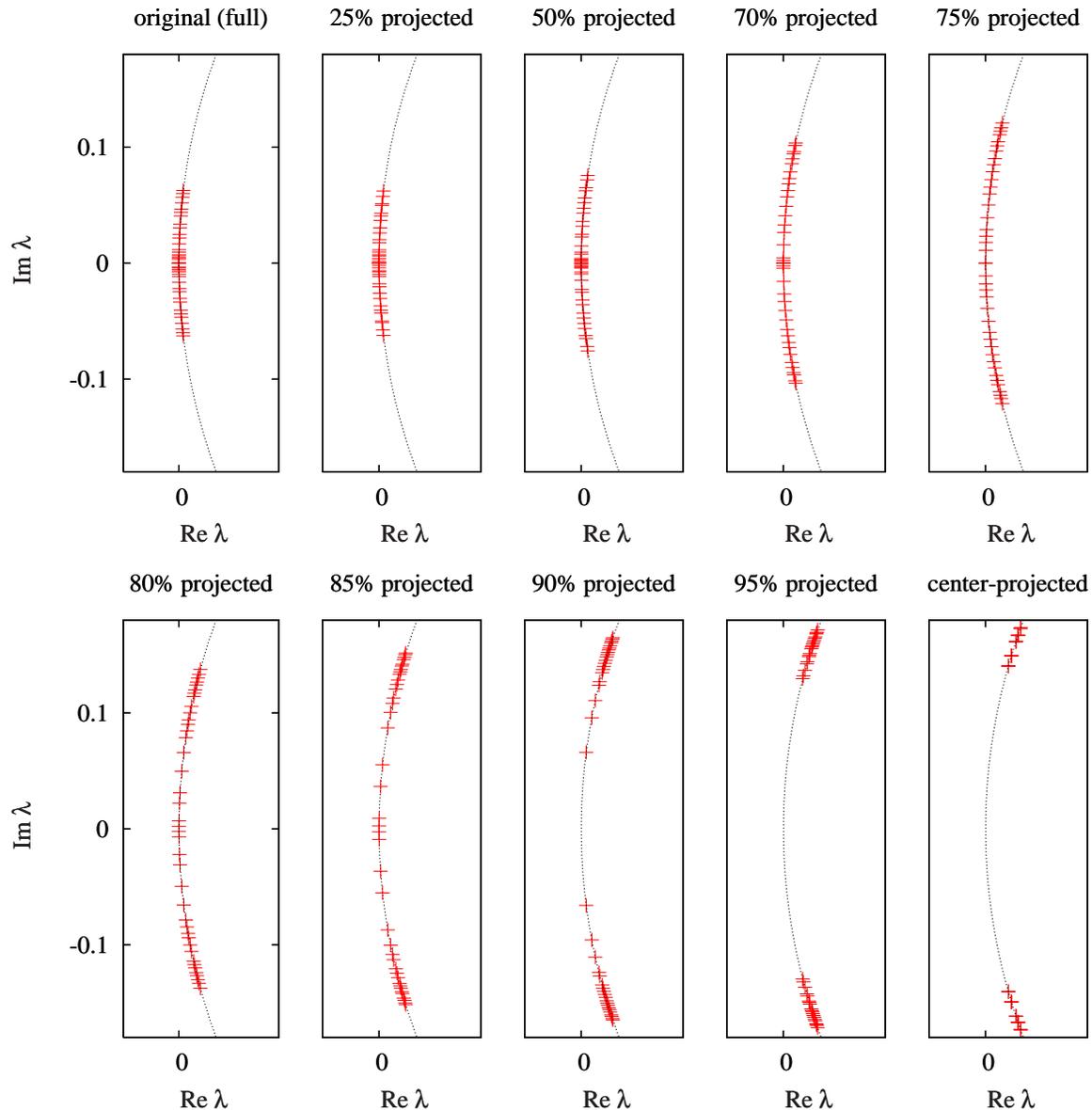}
  \caption{The first twenty overlap Dirac eigenvalue pairs from a single
  configuration on a $16^4$ lattice, antiperiodic boundary conditions at $\b_{LW}=3.3$,  for interpolated fields.}\label{fig:interevs}
\end{figure*}

\begin{figure*}[]
  \centering
  \psfrag{ReL}{Re $\lambda$}
  \psfrag{12x2}{$12^3\times2$}
  \psfrag{12x4}{$12^3\times4$}
  \psfrag{12x6}{$12^3\times6$}
  \psfrag{12x12}{$12^4$}
  \psfrag{ImL}{Im $\lambda$}
  \includegraphics[scale=0.80]{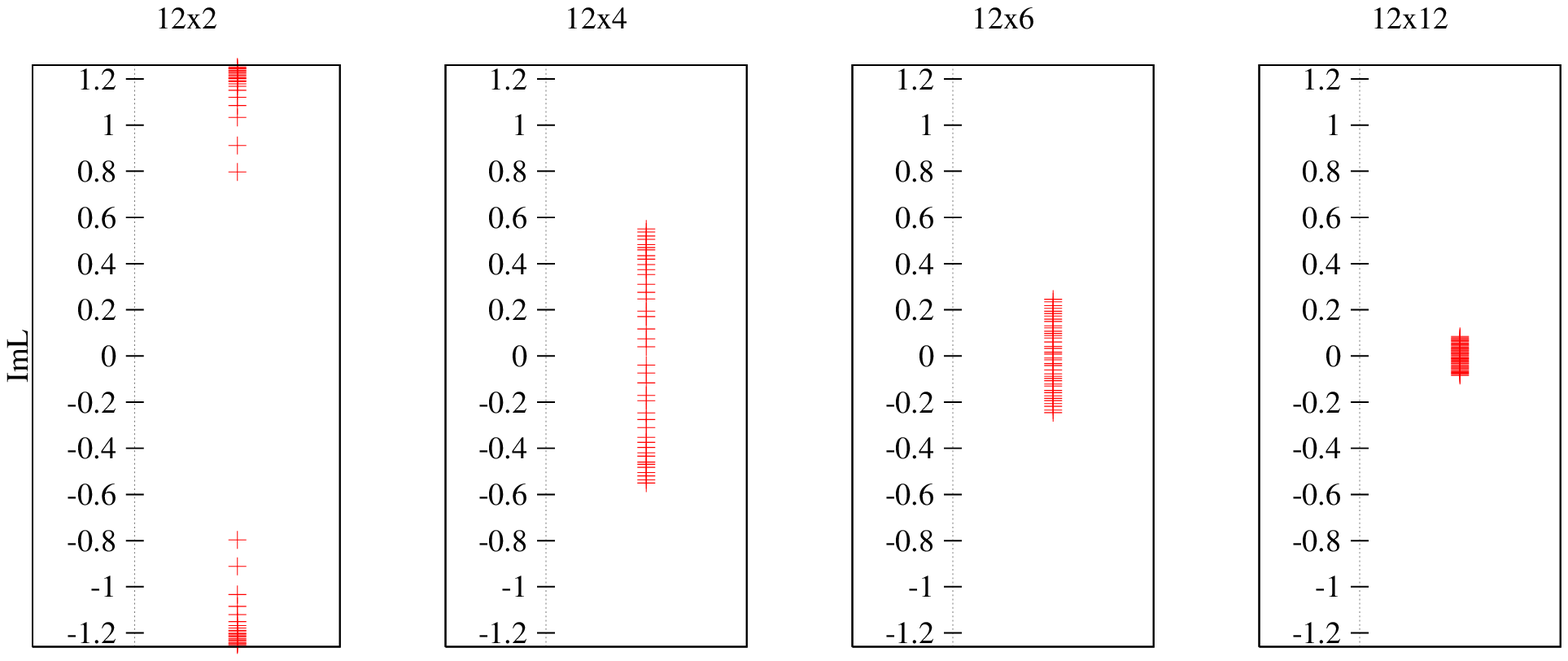}
  \caption{Finite temperature and center projection.  The first twenty asqtad Dirac eigenvalue pairs from $12^3 \times N_T$ 
  center-projected lattices at $\b_{LW}=3.5$ and
  $N_T=2,4,6,12$ lattice spacings, with the usual antiperiodic boundary conditions in the time direction.} 
  \label{highT}
\end{figure*}

In Fig.\ \ref{fig:stevs} we show our results for the twenty lowest-lying eigenvalue pairs of the asqtad Dirac operator, evaluated for periodic and antiperiodic boundary conditions respectively. We see that the eigenvalue gap in the overlap spectrum, found on vortex-only lattices, is erased when eigenvalues are computed for the asqtad Dirac operator.  In fact, the near-zero eigenvalue density in the vortex-only case appears to be substantially enhanced, as compared to the density for unmodified lattices. In the vortex-removed case there also appear to be some near-zero modes centered at $\l=0$, which are separated by a gap from the higher modes.  However, a count of the number of these modes reveals that there are eight doubly degenerate eigenvalues in the central band with Im$(\l) > 0$, and an equal number of complex conjugates with Im$(\l) < 0$, for a total of 32 eigenmodes. Now, the free-field Dirac operator for massless staggered fermions has exactly four zero modes for each of four  ``tastes'', and this number must be multiplied by the number of colors (i.e.\ two for $SU(2)$), for a total of 32 free-field zero modes.  So it is reasonable to guess that the 32 eigenmodes in the central band of the vortex-removed spectrum are simply the would-be zero modes of the free staggered theory. To check this, we carry out the same eigenmode calculation using antiperiodic boundary conditions in one direction, which is sufficient to remove the zero modes of the free theory.   The result is shown in Fig.\ \ref{fig:apstevs}, where it is seen that the boundary conditions have no noticeable effect on the eigenvalue distribution for the  original and center-projected lattices, while the 32 eigenvalues of the central band completely disappear in the vortex-removed case.  This result confirms the conjecture that the central eigenvalues for vortex-removed configurations are simply a remnant of the free-field zero modes, and play no role in \cb. Thus, for the asqtad operator, we have found exactly what was expected prior to the results of Gattnar et al.:  the vortex excitations of the vortex-only lattice carry not only the information about confinement, but are also responsible for \cb~ via the Banks-Casher relation. 
This result was anticipated in ref.\ \cite{AdFE}, which found a non-zero $\langle \overline{\psi} \psi \rangle$ condensate on center-projected lattices.  \cb ~disappears for vortex-removed lattices, as discovered long ago by de Forcrand and D'Elia, in a direct calculation of $\langle \overline{\psi} \psi \rangle$ \cite{dFE}.

If the overlap operator yields misleading results on center-projected lattices, because of the lack of smoothness of center-projected configurations, then perhaps the overlap operator would produce a more reasonable answer when applied to a smoother version of the center-projected lattice.  We therefore consider the following procedure: Given that $SU(2)$ group elements can be represented by unit 4-vectors $a_\m$, where $U=a_0 I_2 + ia_k \s^k$, let $\theta_\m(x)$ denote the angle between the vector representing group element $U_\m(x)$ in maximal center gauge, and the vector representing the $SU(2)$ center element $Z_\m(x) I_2$, where $Z_\m(x)$ was defined in eq.\ \rf{project}. Center projection simply takes this angle to zero, at every link, but we may also consider partial projections in which $\theta_\m(x)$ is everywhere reduced by some fixed percentage. These partial projections interpolate between the unprojected lattice, in maximal center gauge, and the fully center-projected lattice. In Fig.\  \ref{fig:interevs} we show the low-lying eigenvalues for partial projections, with $\theta_\m(x)$ reduced by $25\%,50\%,70\%,75\%,\ldots$, together with the unprojected ($0\%$) and fully ($100\%$) center-projected lattices.  We see that there is no really obvious gap in the partially-projected lattices, even at $80\%$ projection. This agrees with our conjecture that applying the overlap operator to a smoother version of the vortex-only vacuum would give a result consistent with \cb ~and the Banks-Casher relation.

     We conclude this section with a high-temperature result.  Chiral symmetry is restored at high temperatures, and this fact should also hold for center-projected lattices.   Therefore, at sufficiently high-T, a gap should open in the eigenvalue spectrum.  This can be seen in Fig.\ \ref{highT},
where we display the low-lying eigenvalues at $\b_{LW}=3.5$ for time-extensions $N_T=2,4,6,12$ computed on center-projected lattices.  We note that the theory is certainly in the deconfined phase at $N_T=6$, where there is, however, no obvious gap in the eigenvalue spectrum, so it may be that on the projected lattice the chiral transition occurs at a higher temperature than the deconfinement transition.  This is also consistent with ref.\ \cite{AdFE}, which found a nonvanishing  $\overline{\psi}\psi$ condensate at a temperature somewhat above the Wilson action deconfinement temperature.  We should stress, however,
that there is no reason that the chiral and deconfinement temperatures need coincide on the center-projected lattice.   Confinement is a sufficient but not a \emph{necessary} condition for \cb, and, while the center-projected lattice is expected to get the static quark potential about right asymptotically,  this fact certainly does not hold true at intermediate scales, where the finite thickness of real vortices is crucial.  If the static potential on the center-projected lattice is strong enough, \cb ~will be realized, and this symmetry breaking may persist somewhat above the actual deconfinement temperature.  The point is that if we view Yang-Mills configurations as being in some sense factorizable into vortices $\times$ perturbative fluctuations at short distances, then the thickness and internal structure of vortices is important for certain non-perturbative phenomena (such as Casimir scaling, and perhaps the precise chiral transition point) which are sensitive to the static quark potential at intermediate scales.\footnote{Cf.\ ref. \cite{tubby} for a discussion of vortex thickness and Casimir scaling.}

\section{Vortex surfaces and Dirac eigenmode densities}

The breaking of axial U(1) symmetry is associated with topological charge density in the vacuum state, and it is clearly of interest to understand the sources of this topological charge. There is a vast literature on this subject, and candidate sources include instantons, calorons, and intertwined 3-manifolds, with lattice-scale separation, of opposite topological charge density \cite{Ivan}. It has also been suggested, by Engelhardt and Reinhardt \cite{ER}, that topological charge density is concentrated in certain regions of center vortices, where the vortices intersect or ``writhe'' (twist about) in some way.  Since it is generally expected that zero modes of the Dirac operator tend to concentrate in regions of large topological charge density, a correlation between the locations of vortex intersection/writhing points, and the density $\rho_\l(x) = |\psi_\l(x)|^2$ of eigenmodes of the Dirac operator $D$, where $D\psi_\l = \l \psi_\l$ with $\l=0$ (overlap formulation) or $\l \approx 0$ (asqtad), would tend to support the Engelhardt-Reinhardt picture.  A lack of correlation (or perhaps even an anti-correlation) would of course disfavor that picture.

In our investigation we follow the approach of Kovalenko et al.\ in Ref.\ \cite{ITEP}, who worked with eigenmodes of the overlap operator, derived from lattices generated with the Wilson action. Kovalenko et al.\  proposed, as a measure of vortex-eigenmode correlation, the observable 
\beq
C_\l(N_{\rm v})  = \frac{\sum_{p_i}\sum_{x\in H}(V\rho_\lambda(x)-1)}{\sum_{p_i}\sum_{x\in H}1}
\label{correlator}
\eeq
This choice of correlator requires some explanation. Center vortices on the full lattice are located, as explained previously, by center projection in maximal center gauge. Plaquettes on the projected lattice are either $+1$ or $-1$; plaquettes with the latter value are known as ``P-plaquettes''.  However, the thin vortices of the center-projected configurations actually live on the dual lattice. In $D=4$ dimensions each P-plaquette corresponds to a certain plaquette on the dual lattice, and these sets of plaquettes on the dual lattice form closed surfaces, namely, the thin center vortices.  Each point on the vortex surface may belong to a certain number, $N_{\rm v} \ge 3$, of plaquettes on the vortex surface. If the surface is flat at the given point, then $N_{\rm v}=4$. If the point is a corner of the surface, then $N_{\rm v}=3$.  When two flat vortex surfaces intersect at a point (or the same surface intersects itself), then $N_{\rm v}=8$.   It is also possible for the surface to twist (or ``writhe'') around a given point, in such a way that $N_{\rm v}=6$. These writhing points are best visualized by consulting the illustrations in Ref.\ \cite{ER}.  $N_{\rm v}$ can take on other values as well, but $N_{\rm v} =1$ or 2 is impossible for a closed vortex manifold. $N_{\rm v}=0$ holds for all points which do not belong to a vortex surface. For thin vortices, intersection and writhing points are clearly points where the topological charge density is non-zero, and the Engelhardt-Reinhardt proposal is that topological charge is also concentrated on the unprojected lattice in the neighborhood of these locations. In the definition \rf{correlator} of the vortex-eigenmode correlator, the first sum is over all points $p_i$ on the dual lattice which belong to $N_{\rm v}$ plaquettes on the thin vortex surface. The second sum over $x\in H$ is a sum over the 16 points in a hypercube $H$ on the original lattice, surrounding the given point $p_i$ on the dual lattice. $V$ is the lattice volume.   

$C_\l(N_{\rm v})$ is a measure of the average fractional excess of the eigenmode density, $\rho_\l$, over its average value $\langle \rho_\l \rangle = 1/V$, in the neighborhood of a point on the thin vortex connected to $N_{\rm v}$ vortex plaquettes. If this value is non-zero, then the question is whether that value is large enough to be significant. That is not a simple question to answer, in view of the fact that the location of thin vortices varies considerably, from one Gribov copy to another, so we certainly cannot determine the vortex intersections of thick vortices with any great accuracy. This imprecision necessarily lowers the measured valued of the correlator $C_\l(N_{\rm v})$. The best we can do, at this point, is to compare the values of $C_\l(N_{\rm v})$, with Dirac eigenmodes computed on the full lattice, to corresponding values of $C_\l(N_{\rm v})$, with Dirac eigenmodes computed on center-projected lattices.  In the latter case, the only possible source of topological charge is the thin vortices,  the topological charge density is highly localized, and the vortex location is certain.

\begin{figure*}[tbh]
\begin{center}
\subfigure[~Unprojected lattices.]  
{
  \label{va}
  \psfrag{vortex correlation}[0][-1][1][0]{\small vortex correlation}
  \psfrag{number of attached plaquettes}[-1][0][1][0]{\small number of attached plaquettes}
  \psfrag{1st mode}[-1][0][.8][0]{\footnotesize $1^{st}$ mode}
  \psfrag{20th mode}[-1][0][.8][0]{\footnotesize $20^{th}$ mode}
  \includegraphics[scale=0.6]{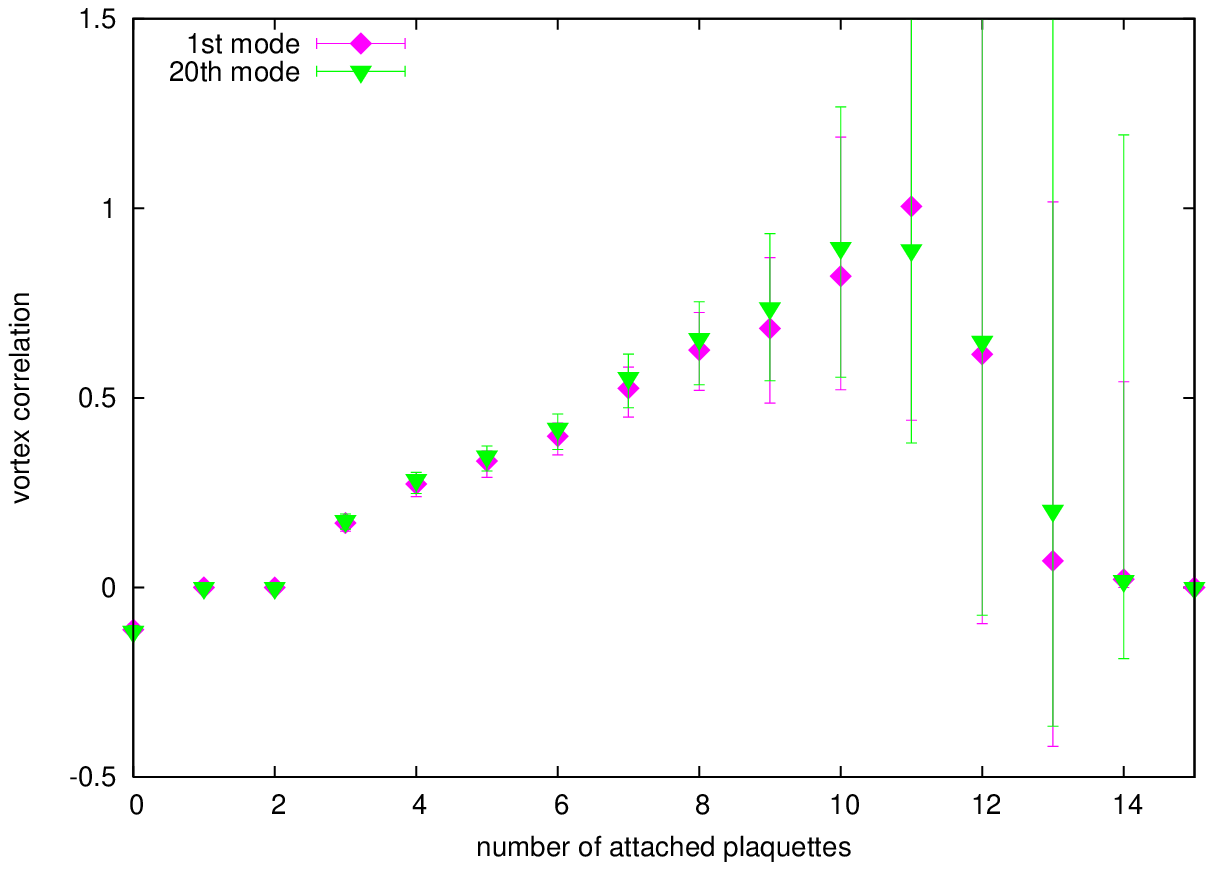}
}
\hspace{0.25cm}
\subfigure[~Center-projected lattices.] 
{
  \label{vb}
  \psfrag{vortex correlation}[0][-1][1][0]{\small vortex correlation}
  \psfrag{number of attached plaquettes}[-1][0][1][0]{\small number of attached plaquettes}
  \psfrag{1st mode}[-1][0][.8][0]{\footnotesize $1^{st}$ mode}
  \psfrag{20th mode}[-1][0][.8][0]{\footnotesize $20^{th}$ mode}
  \includegraphics[scale=0.6]{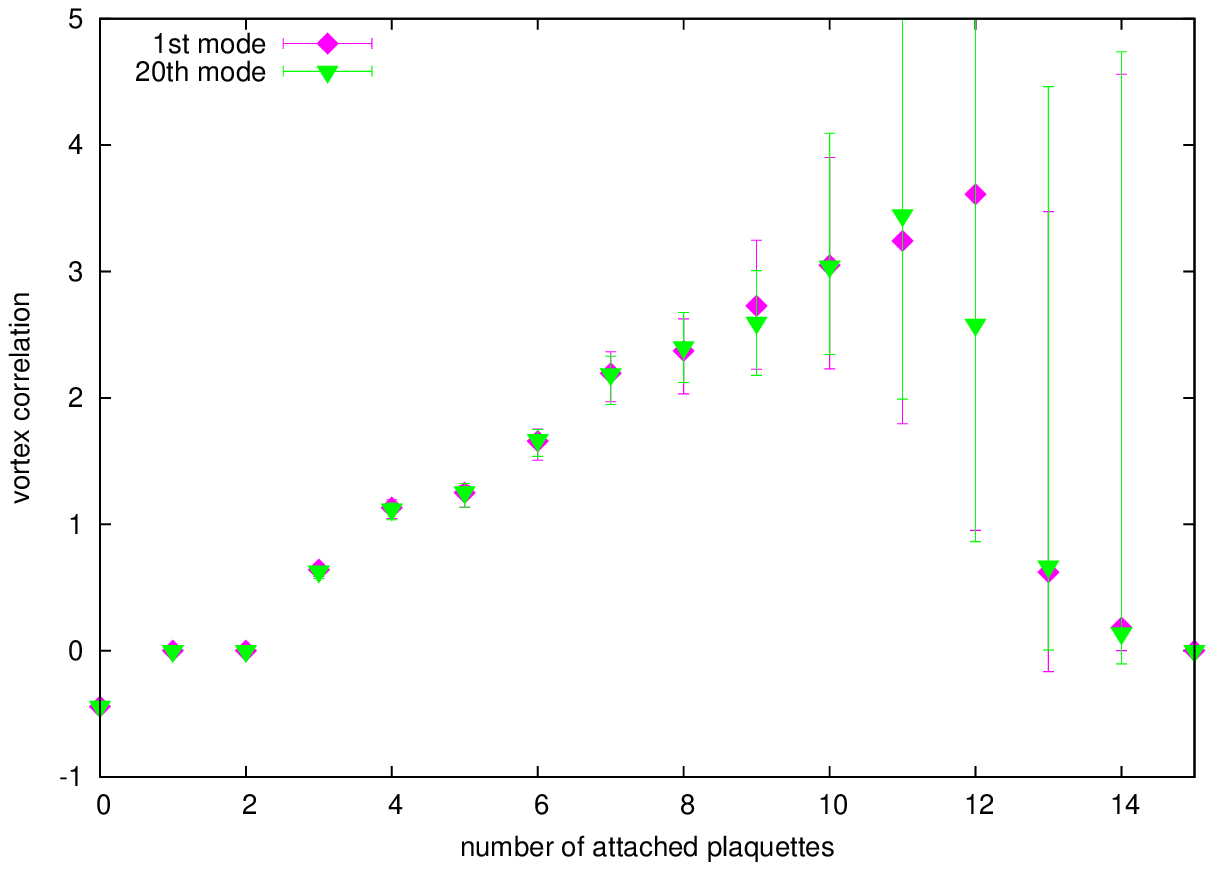}
}
\end{center}
  \caption{Vortex correlation $C_\l(N_{\rm v})$ for asqtad staggered eigenmodes on a $20^4$ lattice at $\b_{LW}=3.3$, 
  a) full and b) center-projected configurations.}
\label{vcasqtad} 
\end{figure*}

\begin{figure*}[htb]  
\begin{center}
\subfigure[~Unprojected lattices.]  
{
  \psfrag{vortex correlation}[0][-1][1][0]{\small vortex correlation}
  \psfrag{number of attached plaquettes}[-1][0][1][0]{\small number of attached plaquettes}
  \psfrag{zero mode}[-1][0][.8][0]{\footnotesize zero mode}
  \psfrag{1st mode}[-1][0][.8][0]{\footnotesize $1^{st}$ mode}
  \psfrag{20th mode}[-1][0][.8][0]{\footnotesize $20^{th}$ mode} 
  \includegraphics[scale=0.6]{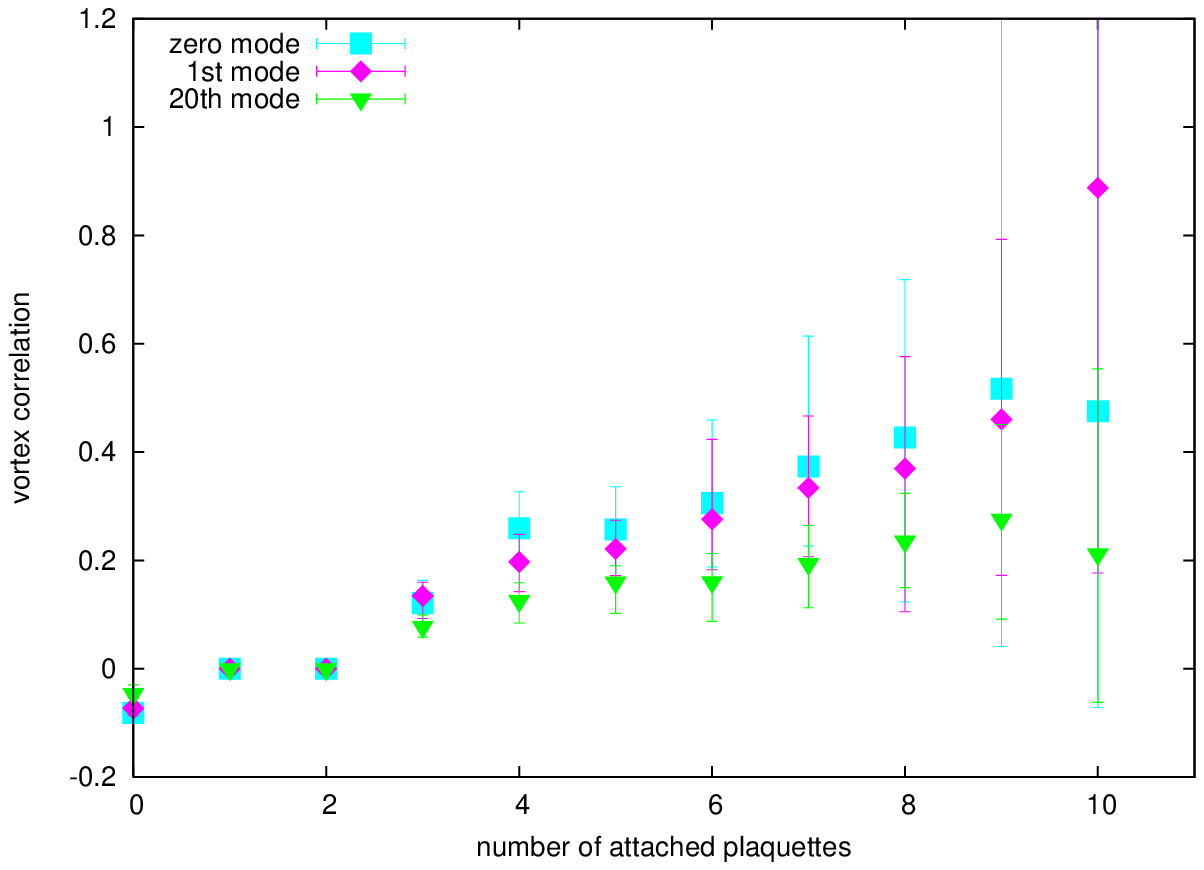}
}
\hspace{0.25cm}
\subfigure[~Center-projected lattices.] 
{
  \psfrag{vortex correlation}[0][-1][1][0]{\small vortex correlation}
  \psfrag{number of attached plaquettes}[-1][0][1][0]{\small number of attached plaquettes}
  \psfrag{zero mode}[-1][0][.8][0]{\footnotesize zero mode}
  \psfrag{1st mode}[-1][0][.8][0]{\footnotesize $1^{st}$ mode}
  \psfrag{20th mode}[-1][0][.8][0]{\footnotesize $20^{th}$ mode}  
  \includegraphics[scale=0.6]{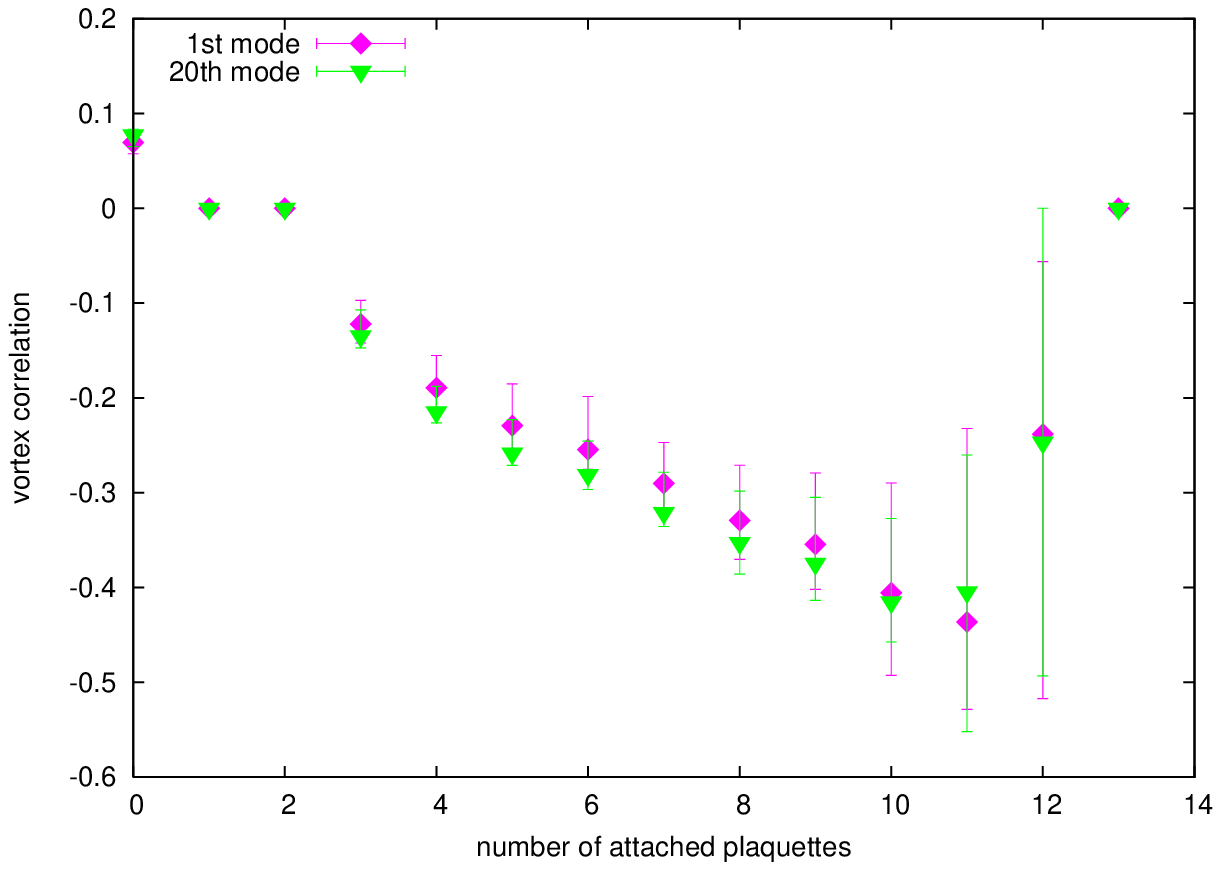}
}
\end{center}
\caption{Vortex correlation $C_\l(N_{\rm v})$ for overlap eigenmodes on a $16^4$ lattice at $\b_{LW}=3.3$, 
a) full and b) center-projected configurations.}
\label{vcovltad}
\end{figure*}

In Fig.\ \ref{vcasqtad} we display the data for $C_\l(N_{\rm v})$ vs.\ $N_{\rm v}$ computed for eigenmodes of the asqtad Dirac operator in the full and (for comparison) center-projected configurations. The lattices are the same as in the previous section, generated by Monte Carlo simulation of the L\"uscher-Weisz action at $\b_{LW}=3.3$. Each plot displays the results for the first eigenmode (lowest $\l$) and the twentieth Dirac eigenmode; we see that for the asqtad eigenmodes there is not much difference.  We find that the values of $C_\l(N_{\rm v})$ obtained from eigenmodes in the full configurations are only about a factor of four smaller than the corresponding values in the center-projected configurations, which in our opinion is not such a great suppression considering (i) the uncertainties in vortex location in unprojected configurations, i.e.\ intersection points of thin vortices may not precisely correspond to the actual intersections of thick center vortices; and (ii) the fact that center vortices are thin, i.e.\ extremely localized, in the center-projected configurations, and therefore one expects a far greater degree of localization in the corresponding Dirac eigenmodes, computed on center-projected lattices.   
But surprisingly, apart from the overall factor of four or so, the Figs. \ref{vcasqtad}a and \ref{vcasqtad}b look much the same. The most important feature, in our opinion, is the fact that the correlator increases steadily with increasing $N_{\rm v}$, and therefore the Dirac eigenmode density is significantly enhanced in regions of large $N_{\rm v}$.  This fact seems at least compatible with the general picture advanced by Engelhardt and Reinhardt. We note, however, that the enhancement seems to be equal for all the low-lying eigenmodes, and not just the ``would-be'' zero-modes of the asqtad operator.

Figure \ref{vcovltad} shows our corresponding results for $C_\l(N_{\rm v})$ vs.\ $N_{\rm v}$, this time computed for eigenmodes of the overlap operator. Our results in Fig.\ \ref{vcovltad}a are consistent with the previous results reported by Kovalenko et al.\ for overlap eigenmodes in Ref.\ \cite{ITEP}.  In the case of the overlap Dirac operator there are true zero modes, and the correlation of the densities of these eigenmodes with vortices is significantly larger than the correlations of the 20th mode. In the case of the overlap operator it is pointless to compute eigenmodes on the center-projected lattice, for reasons we have discussed in previous sections.  If we do this anyway, then we actually find an \emph{anti}correlation between the low-lying modes and  the vortex surfaces, as shown in Fig.\ \ref{vcovltad}b.

\begin{figure*}[tbh]
\begin{center}
\subfigure[~asqtad full lattice]  
{   
 \label{stag-a}
 \includegraphics[scale=0.40]{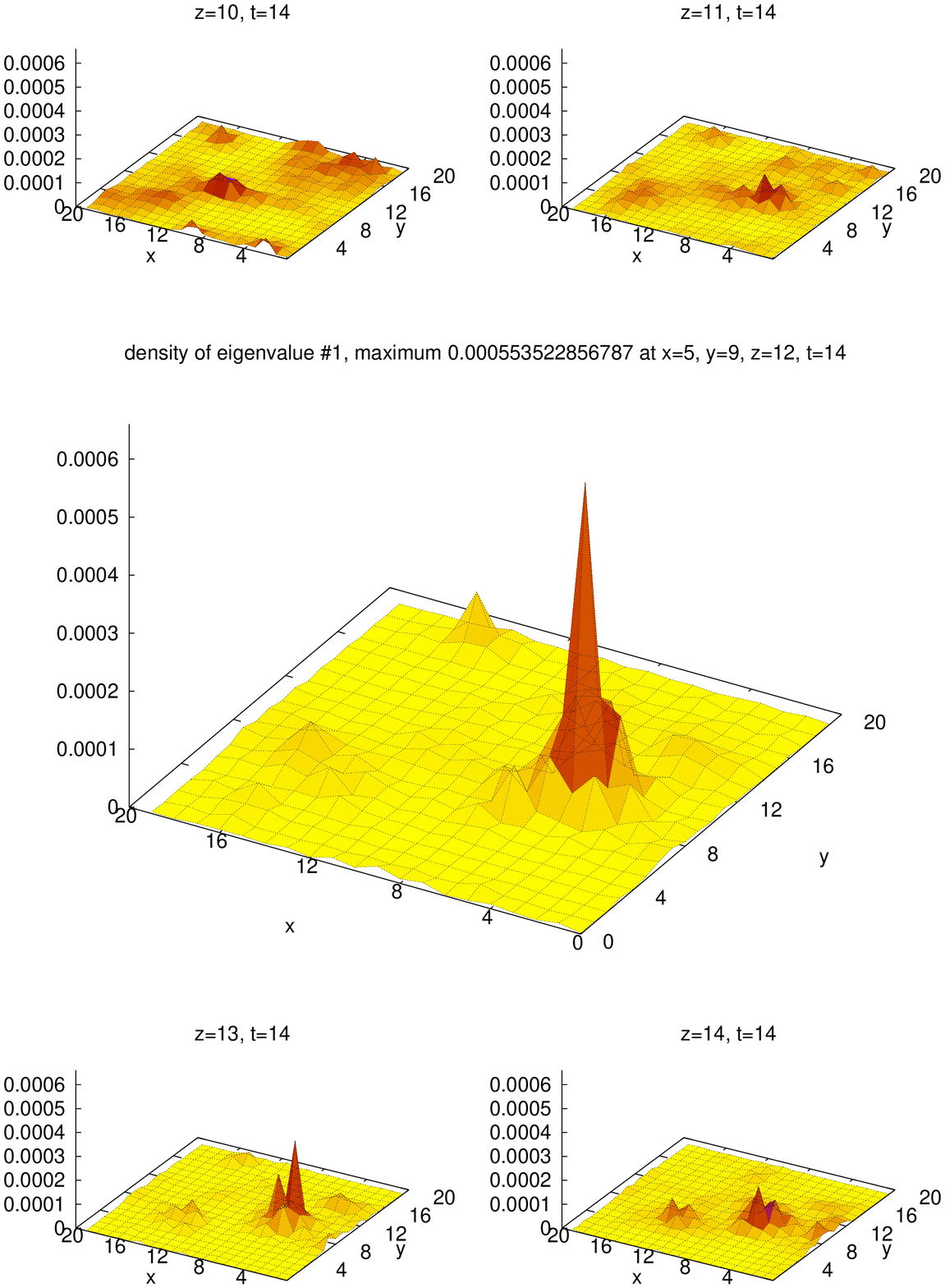}
}
\hspace{0.01cm}
\subfigure[~asqtad center-projected lattice]   
{  
 \label{stag-b}
 \includegraphics[scale=0.40]{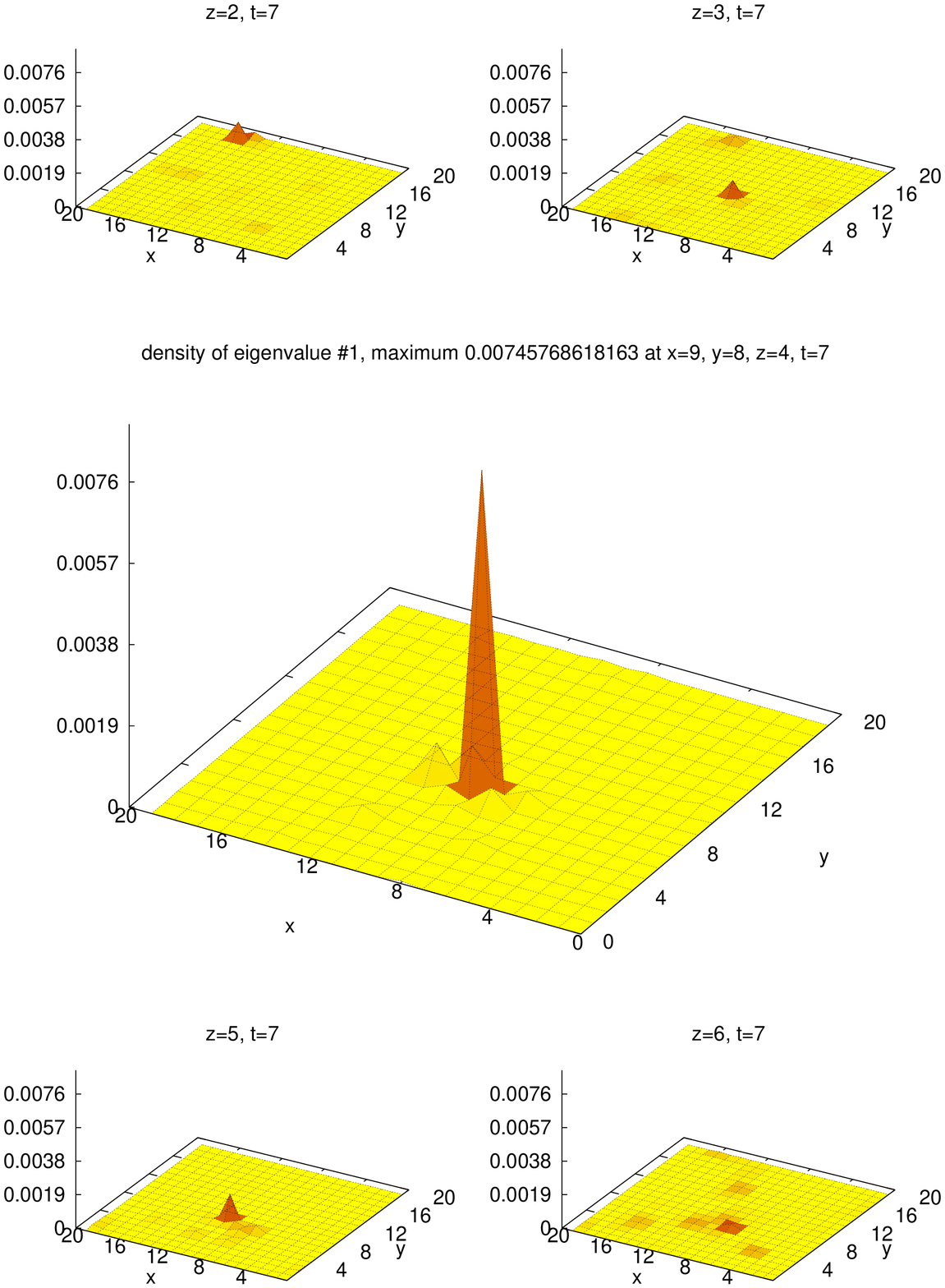}
}
\end{center}
\caption{Maximum density peak (center) of the first \textbf{asqtad} eigenmode on a $20^4$-lattice at $\b_{LW}=3.3$ with upper (above) and lower (below) z-slices of the same t-slice. Eigenmodes are computed on (a) full lattices, and (b) center-projected lattices.}
\label{asqtad_peak} 
\end{figure*}

\section{Dimensionality of Dirac eigenmode concentrations}

\begin{figure*}[t!]
\begin{center}
\subfigure[~overlap full lattice]  
{   
 \label{olap-a}
 \includegraphics[scale=0.40]{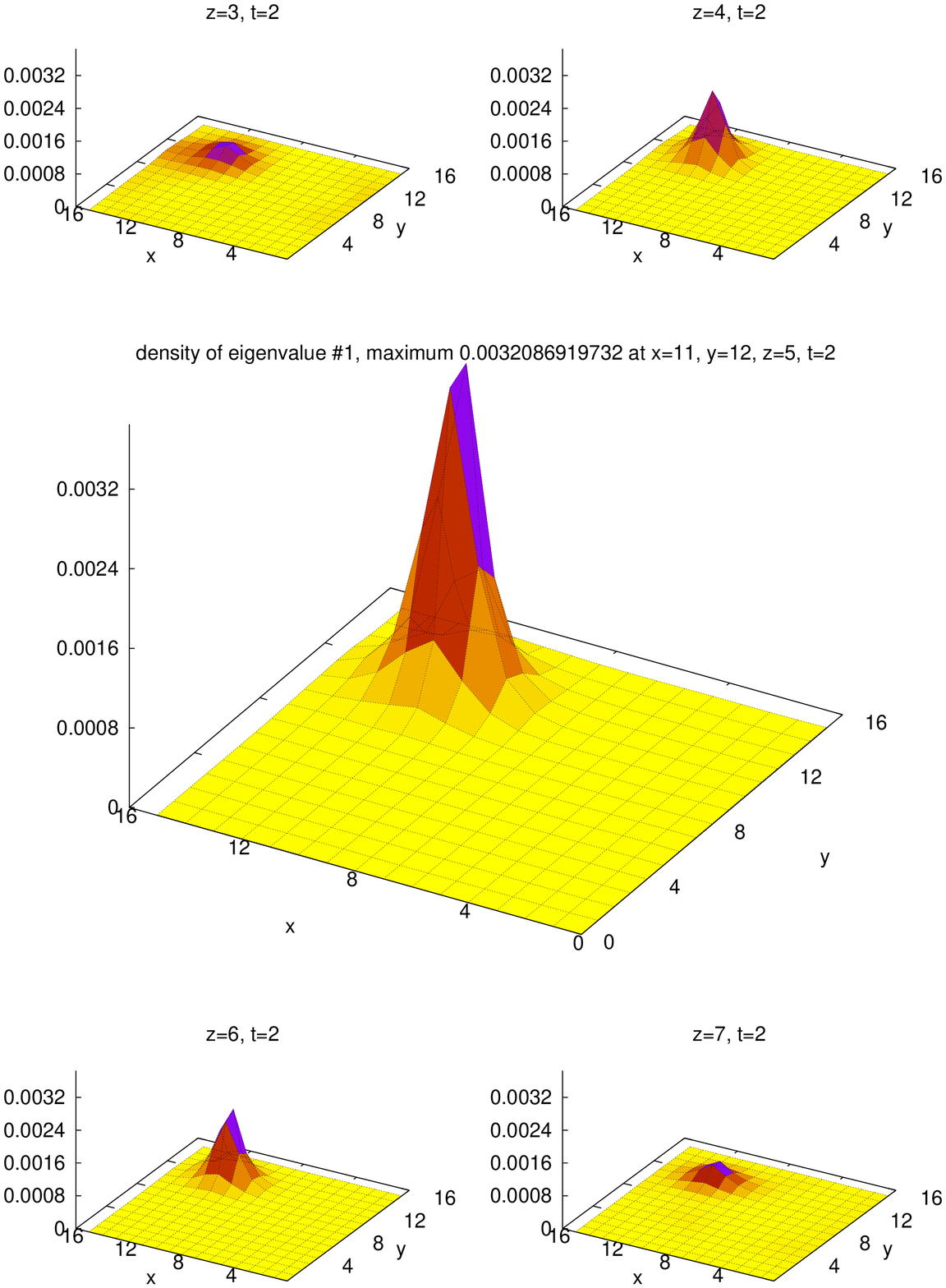}
}
\hspace{0.01cm}
\subfigure[~overlap center-projected lattice]   
{  
 \label{olap-b}
 \includegraphics[scale=0.40]{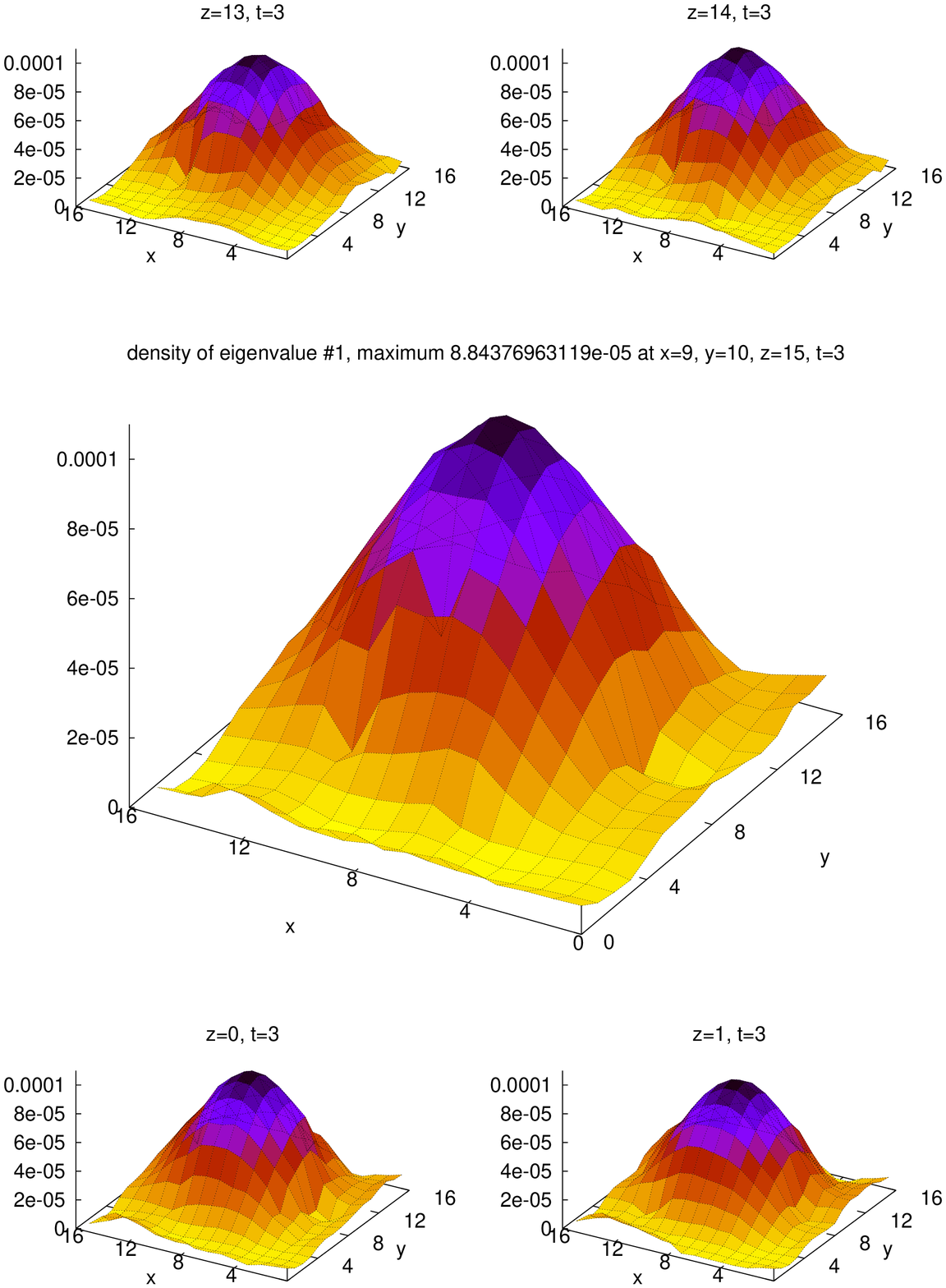}
}
\end{center}
\caption{Maximum density peak (center) of the first \textbf{overlap} eigenmode on a $16^4$-lattice at $\b_{LW}=3.3$ with upper (above) and lower (below) z-slices of the same t-slice. Eigenmodes are computed on (a) full lattices, and (b) center-projected lattices (notice different scales!).}
\label{olap_peak} 
\end{figure*}

The data of the previous section provides some degree of evidence that low-lying Dirac eigenmodes concentrate preferentially at regions on the center vortex surface where there are, e.g., vortex self-intersections, and/or some sort of vortex twisting, such that ``connectivity'' $N_{\rm v}$ of the vortex at a site is larger than $N_{\rm v}=4$. Since surfaces intersect at points in four dimensions, and ``writhing'' points are also points, it is natural to ask whether there is any supporting evidence that the eigenmode density is especially concentrated in point-like regions, rather than along lines, or surfaces, or 3-manifolds.
     
A useful measure to quantify the localization of eigenmodes is the inverse participation ratio (IPR). The IPR of a normalized field $\rho_i(x)$ is defined as
\begin{equation}
   I=N\sum_x\rho_i^2(x)
\end{equation}
where $N$ is the number of lattice sites $x$. Here, $\rho_i(x)=\psi_i^\dag\psi_i(x)$ and $\psi_i(x)$ is the $i$-th, normalized ($\sum_x\rho_i(x)=1$), lowest eigenvector of the Dirac operator. The scaling of the IPR with lattice spacing is sometimes used to determine the dimensionality of eigenmode concentration.  Dimensionality is deduced from the IPR by reasoning that if the eigenmode has support mainly on a submanifold of dimension $d$, with a thickness in the $4-d$ orthogonal directions which is a fixed number of lattice units, then the IPR should scale with lattice spacing as $1/a^{4-d}$.   This reasoning can lead to incorrect conclusions, because it is not necessarily true that the thickness of the localization region is a constant number of lattice spacings, regardless of coupling.

An instructive example is provided by the lowest eigenmode of the covariant Laplacian in the adjoint representation, which was studied in Ref.\ \cite{GORus}. In that case it was found that the IPR scaled like $1/a^2$, suggesting  an eigenmode concentration on surfaces. Instead, it turns out that the lowest eigenmode is sharply concentrated in a pointlike region.  The peculiar scaling of the IPR arises because the volume $b$ of the region of concentration, in lattice units, scales in a peculiar way. If this volume were a constant in physical units, then $ba^4$ would be constant. If instead the volume were constant in lattice units, then $b$ itself would be constant.  In fact, it is $ba^2$ which is constant; the volume of the eigenmode concentration region goes to zero in physical units, but infinity in lattice units, in the continuum limit.  The naive deduction of the dimensionality of the concentration region, purely from the scaling of the IPR, leads in this case to an incorrect conclusion.  For Dirac eigenmodes,  conclusions based on the scaling of the IPR have not been entirely consistent with one another  (cf. the overview in ref.\ \cite{dF-azores}).   Results of the MILC collaboration, with asqtad fermions, indicate a dimensionality $d=3$ \cite{milc}, while the ITEP group has reported results, for overlap fermions and the Wilson action, consistent with $d=0$ \cite{morozov}.  A third study, using overlap fermions and the L\"uscher-Weisz action, again indicates $d=3$ 
\cite{koma} while the latest study of this group, using generalized IPR's defined using higher powers of $\rho_i(x)$, suggest eigenmode concentrations on manifolds of dimension between $d=0$ and $d=1$ \cite{ilgenfritz}. 

A possibly more reliable (if less quantitative) approach is to simply \emph{look} at sample plots of $\rho_\l(x)$ throughout the lattice volume.  What we find, for the eigenmode density of the lowest eigenmodes of the asqtad and overlap operators, is that the eigenmode density is concentrated in very sharp peaks, in point-like regions of the lattice volume.\footnote{The IPR itself is volume independent at fixed lattice spacing, indicating that the number of such peaks is proportional to the lattice volume, as expected.}  In Fig.\ \ref{asqtad_peak}  we display our data for $\rho_\l(x)$, for the lowest eigenmode of the asqtad Dirac operator, in some two-dimensional slices of the four-dimensional lattice volume taken in the neighborhood of the point where $\rho_\l(x)$ is largest.  In Fig.\ \ref{asqtad_peak}(a) we show the density of the lowest eigenmode computed from a typical configuration on the unprojected lattice, and in Fig.\ \ref{asqtad_peak}(b) we show the corresponding data for an eigenmode computed from a typical center-projected lattice.  Each lattice contains several sharp peaks of this kind; it is obvious that the concentration of eigenmode density is in a point-like region, rather than being spread over a submanifold of higher dimensionality. In the figures we display a set of $xy$-plots of $\rho(x,y,z,t)$ at various values of $z$ and fixed $t$, but there is an equally strong falloff of the peak as we move away from the maximum in the time direction. 

Figure \ref{olap_peak}(a) shows the same type of data for a zero mode of the overlap Dirac operator on $16^4$ lattices, again evaluated on unprojected lattices generated from the L\"uscher-Weisz action at $\b_{LW}=3.3$. Here again we find a handful of sharp peaks in the eigenmode density for any thermalized lattice configuration; one such peak is displayed in the plot, and it is concentrated in a point-like region. The situation is very different for eigenmodes of the overlap operator evaluated in center-projected configurations. Instead of having a sharp peak, the eigenmode concentration in this case is very broad, extending over most of the lattice volume, as seen in Fig.\ \ref{olap_peak}(b).\footnote{Note also the considerable difference in vertical scales, between Figs.\  \ref{olap_peak}(a) and \ref{olap_peak}(b).}  However, we have already seen that the overlap operator, evaluated on center-projected configurations, does not have \emph{any} low-lying eigenmodes, let alone a zero mode. It is therefore not surprising that the eigenmode density is qualitatively different from what is found in both the true zero modes of the overlap operator, and the ``would-be'' zero modes of the asqtad operator.

\begin{figure*}[]
\subfigure[~full (unprojected) lattices]
{
  \label{IPR-full}
  \psfrag{l}{\small $\#\lambda$}
  \psfrag{<IPR>}{$\langle\mbox{IPR}\rangle$}
  \psfrag{beta=2.9}[0][0][0.8][0]{\small $\b_{LW}=2.9$}
  \psfrag{beta=3.1}[0][0][0.8][0]{\small $\b_{LW}=3.1$}
  \psfrag{beta=3.3}[0][0][0.8][0]{\small $\b_{LW}=3.3$}
  \psfrag{beta=3.5}[0][0][0.8][0]{\small $\b_{LW}=3.5$}
  \psfrag{beta=3.7}[0][0][0.8][0]{\small $\b_{LW}=3.7$}
  \includegraphics[scale=0.65]{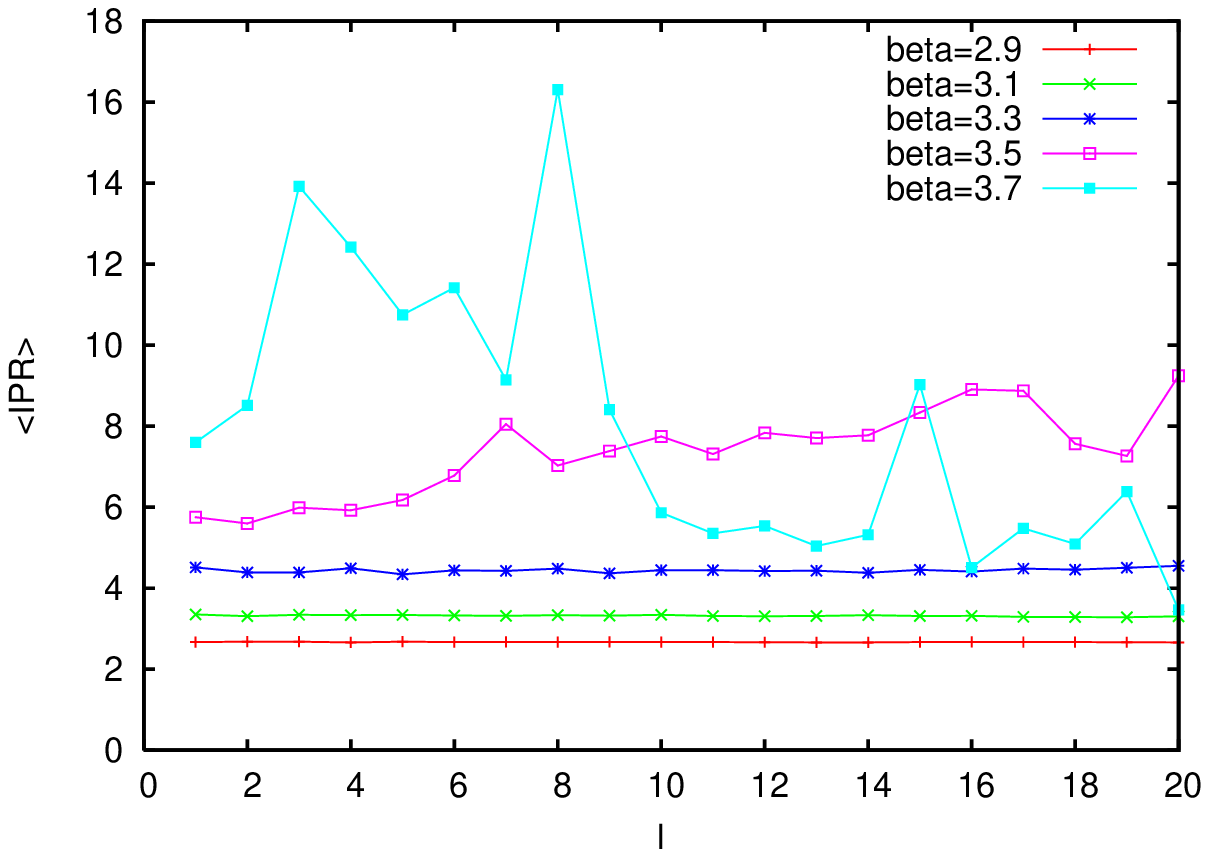}
}
\hspace{0.01cm}
\subfigure[~center-projected lattices]
{ 
  \label{IPR-proj}
  \psfrag{l}{\small $\#\lambda$}
  \psfrag{<IPR>}{$\langle\mbox{IPR}\rangle$}
  \psfrag{beta=2.9}[0][0][0.8][0]{\small $\b_{LW}=2.9$}
  \psfrag{beta=3.1}[0][0][0.8][0]{\small $\b_{LW}=3.1$}
  \psfrag{beta=3.3}[0][0][0.8][0]{\small $\b_{LW}=3.3$}
  \psfrag{beta=3.5}[0][0][0.8][0]{\small $\b_{LW}=3.5$}
  \psfrag{beta=3.7}[0][0][0.8][0]{\small $\b_{LW}=3.7$}
  \includegraphics[scale=0.65]{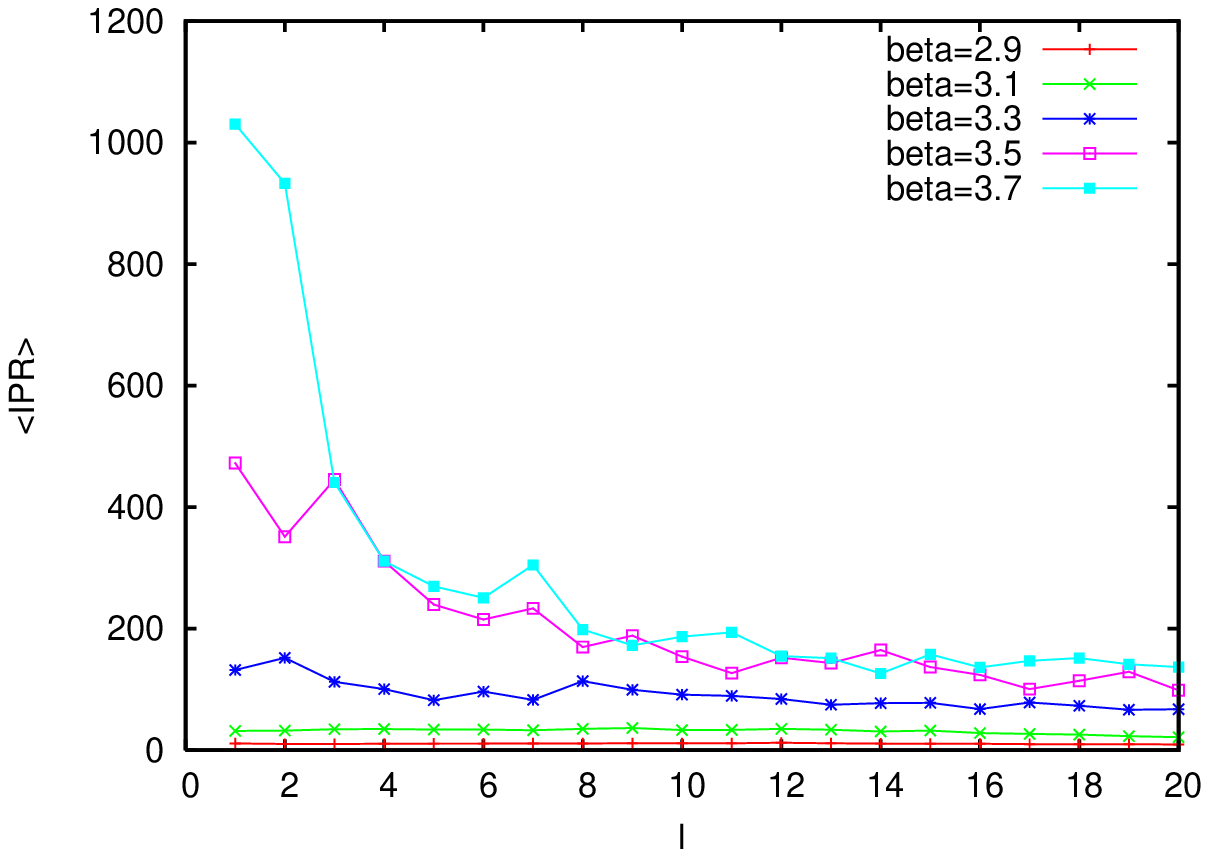} 
}    
\caption{Inverse Participation Ratio vs.\ eigenmode number for asqtad eigenmodes, computed on a $20^4$-lattice at a variety of $\b_{LW}$ values  for: a) unprojected lattices, and b) center-projected lattices.}
\label{IPR}
\end{figure*}

Inspection of Figs.\ \ref{asqtad_peak} and \ref{olap_peak} indicate that eigenmode peaks are by far the sharpest for eigenmodes of the asqtad operator on the center-projected lattice.  Taking account of the vertical scales in these figures, the peak in the asqtad-center-projected case (Fig.\ \ref{asqtad_peak}(b)), is about an order of magnitude higher than the peak in the asqtad-unprojected case (Fig.\ \ref{asqtad_peak}(a)). This difference is of course reflected in a comparison of the IPRs of asqtad eigenmodes on the full and center-projected lattices, shown in Fig.\ \ref{IPR}, which indicate a far higher degree of eigenmode concentration in the center-projected case. 

It is worth noting that on both the unprojected and center-projected lattices, the IPR rises as the lattice spacing becomes smaller, although the rate of increase is seen to be quite different. According to Ref.\ \cite{milc} the IPR should go like $1/a$, in which case it should roughly double in going from $\b_{LW}=2.9$ to $3.3$, $\b_{LW}=3.1$ to $3.5$ or $\b_{LW}=3.3$ to $3.7$.  That behavior is at least crudely compatible with our data (Fig.\ \ref{IPR-full}) for the unprojected lattices. As we have already stressed, this scaling of the IPR does not necessarily mean that eigenmode densities concentrate on three-volumes, and inspection of the density has revealed peaks in the eigenmode density in point-like regions.  For center-projected configurations (Fig.\ref{IPR-proj}) the IPR of the lowest modes is roughly $11$ at $\b_{LW}=2.9$, $33$ at $\b_{LW}=3.1$, $140$ at $\b_{LW}=3.3$, and $400$ at $\b_{LW}=3.5$. Now, if the eigenmode density has support on pointlike regions whose volume is a fixed number of lattice spacings, regardless of $\b_{LW}$, then the IPR should go like $1/a^4$. That means the IPR should increase by about a factor of $4$ from $\b_{LW}=2.9$ to $3.1$, from $\b_{LW}=3.1$ to $3.3$ and from $\b_{LW}=3.3$ to $3.5$. This is not so far off the actual results.

It is not too surprising that would-be zero modes of the asqtad operator would be very highly concentrated when evaluated on center-projected lattices.  On thin vortices, topological charge is concentrated not just in point-like regions, but in fact at individual lattice \emph{sites} on the dual lattice, where thin vortex sheets writhe and/or intersect. Since zero modes concentrate on regions of non-zero topological charge density, and topological charge is concentrated at individual sites on the center-projected lattice, the high degree of localization of the lowest-lying modes, in volumes of lattice-scale extension, is to be expected.  On unprojected lattices the sources of topological charge, whether vortices, instantons, calorons, or something else, are more spread out, and there is no particular reason to expect that the eigenmode density would concentrate in tiny regions of lattice scale extension.  What we would like to know, of course, is which of the candidate sources of topological charge density is giving the main effect.   Our density plots give a strong indication that the charge density concentrates in point-like regions, rather than surfaces or three-volumes, but this fact would be compatible with instanton, caloron, and vortex (intersection/writhing) sources.   The eigenmode density-vortex surface correlations, seen in Figs.\ \ref{vcasqtad} and \ref{vcovltad} provide a modest degree of evidence in support of a vortex origin of topological charge density.  There is also the possibility, of course, that topological charge density may come from more than one type of source.   
 
\section{Conclusions}

We have shown that center-projected $SU(2)$ lattice configurations give rise to a dense low-lying Dirac eigenvalue spectrum, as required for chiral symmetry breaking, for a massless lattice Dirac operator in the asqtad formulation.  In contrast, this low-lying spectrum is \emph{not} found for chirally improved and overlap Dirac operators on center-projected lattices, for reasons which are almost certainly connected to the lack of smoothness of center-projected configurations.   Chiral symmetry is absent in the chirally-improved Dirac operator on such configurations, while for the overlap operator an exact symmetry is present, but is strongly field-dependent for rough configurations (and thus quite different from the continuum symmetry). In the case of the overlap operator we have found that a moderate degree of  ``smoothing'' of the center-projected lattice brings back the low-lying spectrum.  In a staggered formulation such as asqtad, the smoothness of the lattice configuration has nothing to do with the exact chiral symmetry, and the low-lying modes are present for thin vortex configurations. 

There is a general expectation, based on the old Casher argument \cite{Casher}, that gauge field configurations with the confinement property ought to also break chiral symmetry.  Our results indicate that this expectation holds for confining ensembles of thin center vortices, at least when the relevant lattice Dirac operator (asqtad) has some subset of the exact chiral symmetry required by the Casher  argument.  Although non-confining lattice configurations may also break chiral symmetry, we find that vortex-removed configurations (which are non-confining) have a large gap in the asqtad Dirac spectrum around $\l=0$, indicating an unbroken chiral symmetry.

We have also looked at the correlation of vortex location with the densities of low-lying Dirac eigenmodes, following the earlier work of Kovalenko et al.\ \cite{ITEP}. We find that for both asqtad and overlap eigenmodes computed on an unmodified lattice, there is a significant positive correlation between low-lying eigenmode densities and the location of thin vortices on the corresponding center-projected lattice. It is found that this correlation is greatest in the neighborhood of points where a large number of vortex plaquettes meet, such as would be the case for vortex intersections, ``writhings'', or any combination of these effects. Since thin vortices in $D=4$ dimensions intersect and writhe only at points, we would therefore expect that low-lying Dirac eigenmodes are especially concentrated in point-like regions. A simple inspection of eigenmode density distributions reveals that these densities do indeed possess sharp peaks in point-like regions. The vortex-eigenmode correlations, and the peaks in the eigenmode densities at point-like regions,  together provide a degree of support for the picture advanced by Engelhardt and Reinhardt \cite{ER}, in which topological charge would tend to concentrate in the neighborhood of center vortex intersection and writhing points.

It is of interest to compare our results to some related work by Ilgenfritz et al.\ \cite{ilgenfritz}.  These authors report that the high density regions of overlap zero modes are concentrated in regions of dimensionality in the range of zero to one; the lower bound is consistent with our finding (for both overlap and asqtad modes) of sharp peaks in the eigenmode density, located in point-like regions.  We also take note of the earlier work of Gubarev et al.\ in ref.\ \cite{morozov}, which found IPR's for overlap eigenmodes consistent with concentration in point-like regions (for an overview of this and other IPR results, cf.\ \cite{dF-azores}).  Ilgenfritz et al.\  \cite{ilgenfritz} also find a significant positive correlation between the topological charge density at a lattice site, and the probability that a given site is adjacent to a P-vortex sheet or monopole line on the dual lattice.   That result is consistent, assuming a correlation of zero-mode density and topological charge density, with the correlations that we (and Kovalenko et al.\ \cite{ITEP}) have found between vortices and the densities of low-lying Dirac eigenmodes.   

To summarize:  there are significant correlations between center vortices and the low-lying modes of both the asqtad and overlap Dirac operators, and this correlation steadily increases with vortex connectivity ($N_{\rm v}$).   We also find that the thin vortices found in center projection give rise to a low-lying spectrum of Dirac eigenmodes, as required for \cb ~by the Banks-Casher formula, providing that the chiral symmetry of the Dirac operator does not depend on the smoothness of the lattice configuration.  Vortex-removed configurations do not have these low-lying eigenmodes, and therefore do not break chiral symmetry.  Taken together, these results indicate that center vortices have a strong effect on the existence and properties of low-lying eigenmodes of the Dirac operator.

\appendix*
\section{}
 
The gauge action used in this work is a tadpole improved version of the one-loop continuum limit improved $SU(2)$ action of L\"uscher and Weisz~\cite{Luscher:1985zq, Poulis:1997zx}. 

\begin{figure}[h]
\centering
\psfrag{a}{a)}
\psfrag{b}{b)}
\psfrag{c}{c)}
\psfrag{m}{$\mu$}
\psfrag{n}{$\nu$}
\psfrag{l}{$\lambda$}
\includegraphics[scale=0.4]{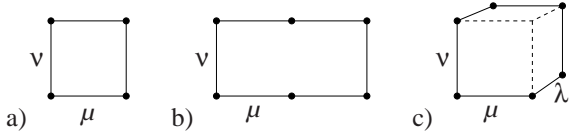}
\caption{\small L\"uscher-Weisz action Wilson loops: a) standard plaquette, b) $2\times1$ rectangle and c) $1\times1\times1$ parallelogram}
\label{fig:lwactloops}
\end{figure}

The standard L\"uscher-Weisz action removes leading ultra-violet cutoff effects by adding a few next-to-nearest neighbour terms to Wilson's action. In addition to the standard plaquette (labeled ``$pl$'') term, it includes a sum over all $2\times1$ (planar) rectangle (labeled ``$rt$'') and over all $1\times1\times1$ parallelogram (labeled ``$pg$'') Wilson loop terms (see Fig.~\ref{fig:lwactloops}). For $SU(N)$ lattice gauge fields $U_\mu(x)$ living on a four-dimensional ($\mu=0,1,2,3$) hypercubic lattice with sites $x$ and lattice spacing $a$, the improved action reads
\begin{equation}
S[U]=\beta\sum_x\left\{c_{pl}\sum_{pl}S_{pl}+c_{rt}\sum_{rt}S_{rt}+c_{pg}\sum_{pg}S_{pg}\right\}
\end{equation}
where $\beta$ denotes the (inverse) coupling constant and $S_i=\frac{1}{N}\mbox{Re Tr}(1-U_i)$ with $U_i$ the corresponding Wilson loops. The coefficients $c_i=c_i^0+4\pi\alpha_0\Delta_i$ for one-loop corrections $\Delta_i$ have been computed by L\"uscher and Weisz for both $SU(2)$ and $SU(3)$ (Table 1 in Ref.~\cite{Luscher:1985zq}).
The continuum limit behavior of the L\"uscher-Weisz action can be further improved by making the lattice links more ``continuum like''. At the mean field level this entails setting $U_\mu\rightarrow u_0^{-1}U_\mu$, where one possible choice for the mean field (or ``tadpole'') factor $u_0$ is using the expectation value of the average plaquette
\begin{equation}
u_0=<\frac{1}{N}\mbox{Re Tr}U_{pl}>^{1/4}.
\end{equation}
The L\"uscher-Weisz action can now be tadpole improved by explicitly pulling a $u_0^{-1}$ factor out of each link and replacing $\alpha_0$ in the one-loop perturbatively renormalized coefficients $c_i$ with a nonperturbatively renormalized coupling $\alpha_s$ defined through~\cite{Poulis:1997zx}
\begin{equation}
  \begin{aligned}\alpha_s&=-4\frac{\ln u_0}{\xi_N}\quad\mbox{with}\\
\xi_N&=0.366262\pi\frac{N^2-1}{N}=\left\{\begin{aligned}&1.72597,\quad\mbox{for~} N=2\\ &3.06839,\quad\mbox{for~} N=3\end{aligned}\right.\end{aligned}
\end{equation}
Defining $\beta_{LW}\equiv u_0^{-4}\beta$ (since $U_{pl}$ involves $4$ links) the improved action reads for $SU(2)$~\cite{Poulis:1997zx}
\begin{equation}
\begin{aligned} S=\beta_{LW}\sum_{pl}S_{pl}&-\frac{\beta_{LW}}{20u_0^2}[1+0.2227\alpha_s]\sum_{rt}S_{rt}\\ &-0.02224\frac{\beta_{LW}}{u_0^2}\alpha_s\sum_{pg}S_{pg}\end{aligned}.
\end{equation}
The tadpole factor $u_0$ is determined during thermalization and then kept fixed. In view of extracting physical quantities, we fit the time-dependent potential $V(R,T)=\log(W[R,T-1]/W[R,T])$ with $W[R,T]$ the Wilson loop of size $R\times T$ in space-/time-direction respectively at some fixed $T$, to an ansatz $V(R)=\sigma_TR-c/R+v_0$ (linear-plus-Coulomb fit). In order to obtain an asymptotic lattice string tension $\sigma_{lat}$ we fit the extracted string tensions $\sigma_T$ for several $T$ values to some stabilizing function $f(T)=\exp(-kT+d)+\sigma$ with $\sigma$ giving the asymptotic ($T\rightarrow\infty$) value. All fits were done by least-square routines. 
To set the scale we use the physical string tension, $\sqrt{\sigma_{lat}}/a=\sqrt{\sigma_{phys}}\approx0.44$GeV~\cite{Bali}, to determine the lattice spacing $a$. Table~\ref{tab:lwdata} lists the data for runs of string tension determination on $20^4$-lattices with a $1000$ thermalization steps, $1000$ measurements separated by $200$ iterations each.

\begin{table}[h!]
  \centering
  \begin{tabular}{c|cc|cc}
    $\beta_{LW}$&$\sigma_{lat}$&&$a$[fm]&\\
    \hline
    $2.9$&$0.3756$&$\pm0.0053$&$0.2749$&$\pm0.0019$\\
    $3.1$&$0.2254$&$\pm0.0033$&$0.2129$&$\pm0.0016$\\
    $3.3$&$0.1112$&$\pm0.0017$&$0.1495$&$\pm0.0012$\\
    $3.5$&$0.0635$&$\pm0.0007$&$0.1138$&$\pm0.0006$\\
    $3.7$&$0.0401$&$\pm0.0003$&$0.0898$&$\pm0.0003$\\
    $4.0$&$0.0225$&$\pm0.0002$&$0.0673$&$\pm0.0002$\\
  \end{tabular}
  \caption{\small Lattice string tension $\sigma_{lat}$ and lattice spacing $a$, as extracted from linear-plus-Coulomb fits to $V(R,T=fixed)$}
  \label{tab:lwdata}
\end{table}

 \begin{figure*}[t!]
 \subfigure[~P-vortex surface density vs.\ coupling constant 
  $\beta_{LW}$.  ``Two loop'' line is the scaling prediction  
  with $\sqrt{\rho/6\Lambda^2}=50$.]  
{   
  \psfrag{4}{}
  \psfrag{blw}{\small $\beta_{LW}$}
  \psfrag{vortex density}[0][-.4][1][0]{\small vortex density}
  \psfrag{two loop}[0][0][.8][0]{\small two loop}
  \psfrag{L=8}[0][0][.8][0]{\small L=$8^4$}
  \psfrag{L=12}[0][0][.8][0]{\small L=$12^4$}
  \psfrag{L=16}[0][0][.8][0]{\small L=$16^4$}
  \psfrag{L=20}[0][0][.8][0]{\small L=$20^4$}
  \centering
  \includegraphics[scale=0.65]{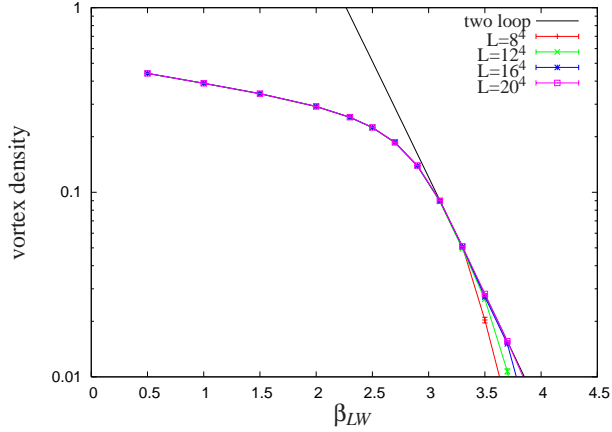}   
   \label{fig:vdenslog}
 }
 \hspace{0.01cm}
\subfigure[~Center-projected Creutz-ratios at $\beta_{LW}=3.1-3.5$. Horizontal bands
   indicate asymptotic string tensions on the unprojected lattice.]   
{    
  \psfrag{b=2.9}[0][0][.8][0]{\small $\b_{LW}=2.9$}
  \psfrag{b=3.1}[0][0][.8][0]{\small $\b_{LW}=3.1$}
  \psfrag{b=3.3}[0][0][.8][0]{\small $\b_{LW}=3.3$}
  \psfrag{b=3.5}[0][0][.8][0]{\small $\b_{LW}=3.5$}
  \psfrag{b=3.7}[0][0][.8][0]{\small $\b_{LW}=3.7$}
  \psfrag{x(R,R)}{\small $\chi(R,R)$}
  \psfrag{asymptotic string tension}[-2][0][1][0]{\footnotesize asympt. string tension}
  \psfrag{R}{\small $R$}
  \psfrag{4}{}
  \psfrag{L=8}[0][0][.8][0]{\small L=$8^4$}
  \psfrag{L=12}[0][0][.8][0]{\small L=$12^4$}
  \psfrag{L=16}[0][0][.8][0]{\small L=$16^4$}
  \psfrag{L=20}[0][0][.8][0]{\small L=$20^4$}
  \includegraphics[scale=0.65]{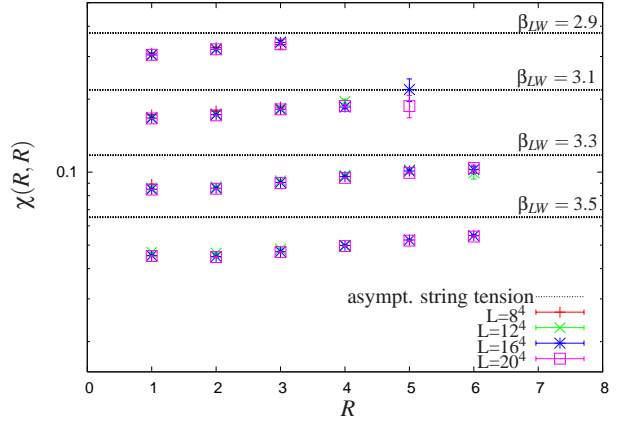} 
  \label{centdom}
}
\caption{Evidence for:  a) asymptotic scaling of vortex densities; and b) center dominance
for the string tension.}
 \end{figure*}
 
\begin{figure*}[htb]
\subfigure[~Vortex limited loops.]  
{  
  \label{Wn} 
  \psfrag{Wn/W0}{\small $W_n/W_0$}
  \psfrag{W1/W0}{\small $W_1/W_0$}
  \psfrag{W2/W0}{\small $W_2/W_0$}
  \psfrag{Loop Area}{\small Loop Area} 
  \includegraphics[scale=0.66]{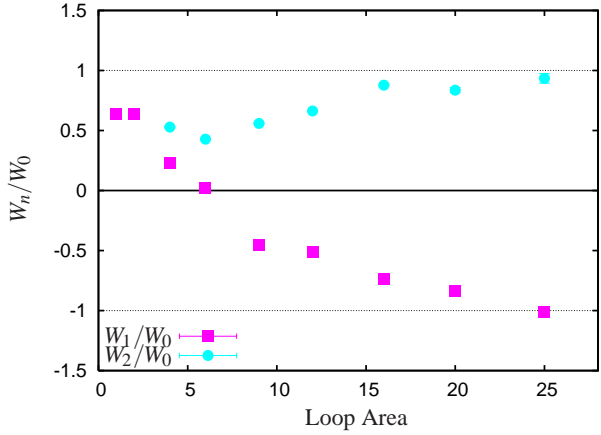}
}
\hspace{0.01cm}
\subfigure[~Vortex removal.]  
{  
  \label{Wrem}
  \psfrag{x(R,R)}{$\chi(R,R)$}
  \psfrag{asymptotic string tension}[-.3][0][1][0]{\footnotesize asymptotic string tension}
  \psfrag{R}{\small $R$}
  \psfrag{full theory}[0][0][1][0]{\footnotesize full theory}
  \psfrag{center projected}[0][0][1][0]{\footnotesize center-projected}
  \psfrag{vortex removed}[0][0][1][0]{\footnotesize vortex-removed}
  \includegraphics[scale=0.66]{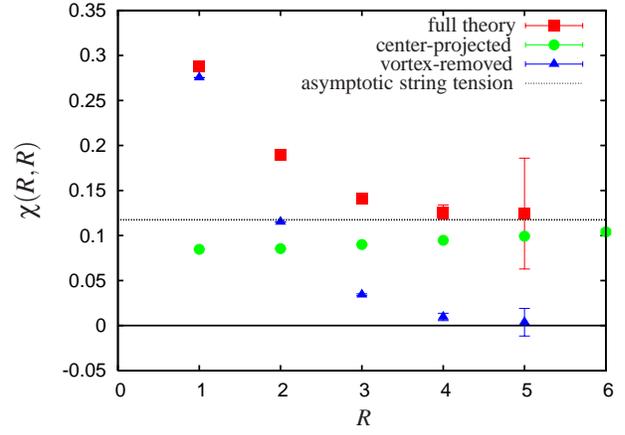}
}
 \caption{Effects of vortex restriction, and vortex removal, on Wilson loops. (a) Vortex limited Wilson loops $W_n$ on $20^4$ lattice at $\beta_{LW}=3.1$.  Note the approach, as loop area increases, to the limit $(-1)^nW_n$.
   (b) The effect of vortex removal. $20^4$ lattice at $\beta_{LW}=3.3$ for full, center-projected and vortex-removed data. Horizontal bands indicate asymptotic string tension on the unprojected lattice with errorbars.}
  \label{W}
\end{figure*}

All previous numerical checks of the vortex mechanism, via maximal center gauge fixing and center projection, have been carried out on thermalized lattices generated from Monte Carlo simulations of the Wilson action (a discussion of these tests and their significance can be found in Ref.\ \cite{review}.) It is important to repeat these checks for the new action. It turns out that there are no surprises, and vortex results derived using the L\"uscher-Weisz action are consistent with the previous work in the Wilson action.   Here the gauge-fixing procedure is the over-relaxation method applied to one configuration, resulting in a single (random) gauge copy.  No attempt is made to find the true global maximum of squared link variables, or the best out of a set, so technically the procedure corresponds to averaging over all gauge copies of direct maximal center gauge in the Gribov region. We have compared this procedure to the ``best copy of five" prescription, and have found no significant difference in the results.    

Figure \ref{fig:vdenslog} shows the P-vortex density (in lattice units) vs.\  $\b_{LW}$. The solid line is the two-loop asymptotic freedom behavior for this quantity, for a choice of vortex density $\rho$ (area per unit volume) in physical units satisfying $\sqrt{\rho/6\Lambda^2}=50$. Figure \ref{centdom} is a test of vortex dominance; the figure shows the values for Creutz ratios $\chi(R,R)$ on the center-projected lattice at various $R$ and various couplings.  The horizontal bands indicate the asymptotic string tension on unprojected lattices (middle lines), together with their corresponding error bars  (upper and lower lines). The center-projected Creutz ratios approach the asymptotic string tension, but at largest available $R$ values are still lower by about $10\%$.

In Fig.\ \ref{Wn} we show our data for the ratios of vortex-limited Wilson loops $W_1/W_0$ and $W_2/W_0$. A vortex-limited Wilson loop is a loop built of unprojected links, which is pierced by $n$ P-vortices on the corresponding center-projected lattices.  It is expected that for sufficiently large loops in $SU(2)$ gauge theory, $W_1/W_0 \ra -1$ and $W_2/W_0 \ra 1$, at least if the vortex piercings are near the middle of the loop, rather than lying near the perimeter. The data shown in Fig.\ \ref{Wn} was collected with this restriction on P-vortex piercings.  Finally, the effect of vortex removal on Creutz ratios is shown in Fig.\ \ref{Wrem}, together with full and center-projected results, at $\b_{LW}=3.3$.   All of this data is similar to corresponding results previously obtained from simulations of the $SU(2)$ Wilson action. \\

\acknowledgments{This research was supported in part by the U.S.\ Department of Energy under Grant No.\ DE-FG03-92ER40711 (J.G.), by the Slovak Grant Agency for Science, Project VEGA No.\ 2/6068/2006 (\v{S}.O.) and by ``Fonds zur F\"orderung der Wissenschaften'' (FWF) under contract P20016-N16 (R.H.).}

\end{document}